\documentclass[a4paper,11pt]{article}
\pdfoutput=1
\usepackage{jcappub} 
\usepackage{natbib}
\setcitestyle{square,comma,numbers,sort&compress}
\usepackage{enumerate}
\usepackage{tabularx}
\usepackage[left = 1.0in, top=1.0in,right=1.0in,bottom=1.0in,a4paper]{geometry}
\usepackage[dvipsnames]{xcolor}
\usepackage{comment}
\usepackage[caption=false]{subfig}
\usepackage{cancel}
\def\mc#1{\mathcal#1}
\def\del {\partial}
\usepackage[section]{placeins}
\usepackage{listings}
\usepackage{amsbsy}
\newcolumntype{Y}{>{\centering\arraybackslash}X}

\definecolor{myRED}{rgb}{0.8, 0.25, 0.33}

\title{\boldmath\huge BBN Photodisintegration Constraints on Gravitationally Produced Vector Bosons}

\author[a]{Chee Sheng Fong,} 
\author[b]{Moinul Hossain Rahat,} 
\author[c]{and Shaikh Saad} 

\affiliation[a]{Centro de Ciências Naturais e Humanas,
Universidade Federal do ABC, 09.210-170, Santo André, SP, Brazil}
\affiliation[b]{Institute for Fundamental Theory, Department of Physics,
University of Florida, Gainesville, FL 32611, USA }
\affiliation[c]{Department of Physics, University of Basel, Klingelbergstr.\ 82, CH-4056 Basel, Switzerland}

\emailAdd{sheng.fong@ufabc.edu.br}
\emailAdd{mrahat@ufl.edu}
\emailAdd{shaikh.saad@unibas.ch}

\abstract{
Gravitational production of massive particles due to cosmic expansion can be significant during the inflationary and reheating period of the Universe. If the particle also has non-gravitational interactions that do not significantly affect its production, numerous observational probes open up, including cosmological probes. In this work, we focus on the gravitational production of light vector bosons that couple feebly to the Standard Model (SM) particles. Due to the very feeble coupling, the light vector bosons never reach thermal equilibrium, and if the Hubble scale at the end of inflation is above $10^8$\,GeV, the gravitational production can overwhelm the thermal production via the freeze-in mechanism by many orders of magnitude. As a result, much stronger constraints from the Big Bang Nucleosynthesis (BBN) can be placed on the lifetime and mass of the vector bosons compared to the scenario where only thermal production is considered. As an example, we study the sub-GeV scale dark photons, which couple to the SM only through kinetic mixing, and derive constraints on the mass and kinetic mixing parameter of the dark photon from the photodisintegration effects on the light element abundances relevant at the end of the BBN when the cosmic age was around $10^4$\,s.
}

\makeatletter
\gdef\@fpheader{}
\makeatother

\begin{document}
\maketitle
\flushbottom

%%%%%%%%%%%%%%%%%%%%%%%%%%%%%%%%%%%%%%%%
%%%%%%%%%%%%%%%%%%%%%%%%%%%%%%%%%%%%%%%%
\section{Introduction}
The possibility of particle production in an expanding Universe was first raised by Schr\"{o}dinger in 1939~\cite{1939Phy.....6..899S} and the calculation for particle creation for quantum fields in the Friedmann-Robertson-Walker background was first carried out by Parker~\cite{Parker:1968mv,Parker:1969au,Parker:1971pt}. While the particle creation is quite negligible at the present time, it can be significant during the inflationary and reheating period of the Universe~\cite{Ford:1986sy,Lyth:1996yj}. Any fields which are not conformal invariant will be produced from cosmic expansion, and hence all massive particles will necessarily be produced gravitationally during the cosmic evolution. 

In this work, we consider light vector bosons in the mass range MeV to GeV that couple feebly to the Standard Model (SM) such that they are never in thermal equilibrium. We take into account their gravitational production during the period of inflation and reheating of the Universe~\cite{Dimopoulos:2006ms,Graham:2015rva,Ema:2019yrd,Ahmed:2020fhc,Kolb:2020fwh}. When the Hubble scale at the end of inflation $H_I$ is greater than $10^8$\,GeV, gravitational production starts to dominate over the thermal production from the freeze-in mechanism, leading to stronger cosmological constraints on the properties of vector bosons.\footnote{Gravitational production from scatterings with the SM fields~\cite{Garny:2015sjg,Tang:2017hvq,Bernal:2018qlk} or inflatons~\cite{Ema:2015dka,Ema:2016hlw,Ema:2018ucl,Mambrini:2021zpp,Barman:2021ugy} are always subdominant for $H_I \gtrsim 10^{8}\,{\rm GeV}$ and will not be considered in this work.} 

For the cosmological constraints, we will focus on the scenario where their lifetime is greater than about $10^4$\,s (thermal bath temperature $T \lesssim 10$\,keV) after the completion of the Big Bang Nucleosynthesis (BBN).
From their decays, the electromagnetic injections (photons and electron/positrons) will induce an electromagnetic cascade through rapid interactions with the background photons and electrons and give rise to a photon spectrum below the energy threshold of pair production $E_{\rm th}^{e^-e^+} \approx \frac{m_e^2}{22T}$. For $T \lesssim 10$\,keV, these photons will have sufficient energy to destroy the light elements leading to the so-called photodisintegration or photodissociation effects. If the electromagnetic injection is smaller than $E_{\rm th}^{e^-e^+}$, the approximation of a universal photon spectrum starts to break down~\cite{Poulin:2015woa,Poulin:2015opa,Hufnagel:2018bjp,Forestell:2018txr}. Hence for the case of light vector boson, we resort to using the public code \texttt{ACROPOLIS}~\cite{Depta:2020mhj,Depta:2020zbh,Hufnagel:2018bjp} to calculate the electromagnetic cascade from the decays of vector bosons and the photodisintegration effects on the light element abundances.

\begin{figure}[!ht] 
    \centering
    \subfloat[\label{moneyplota}]{
        \includegraphics[width=0.49\textwidth]{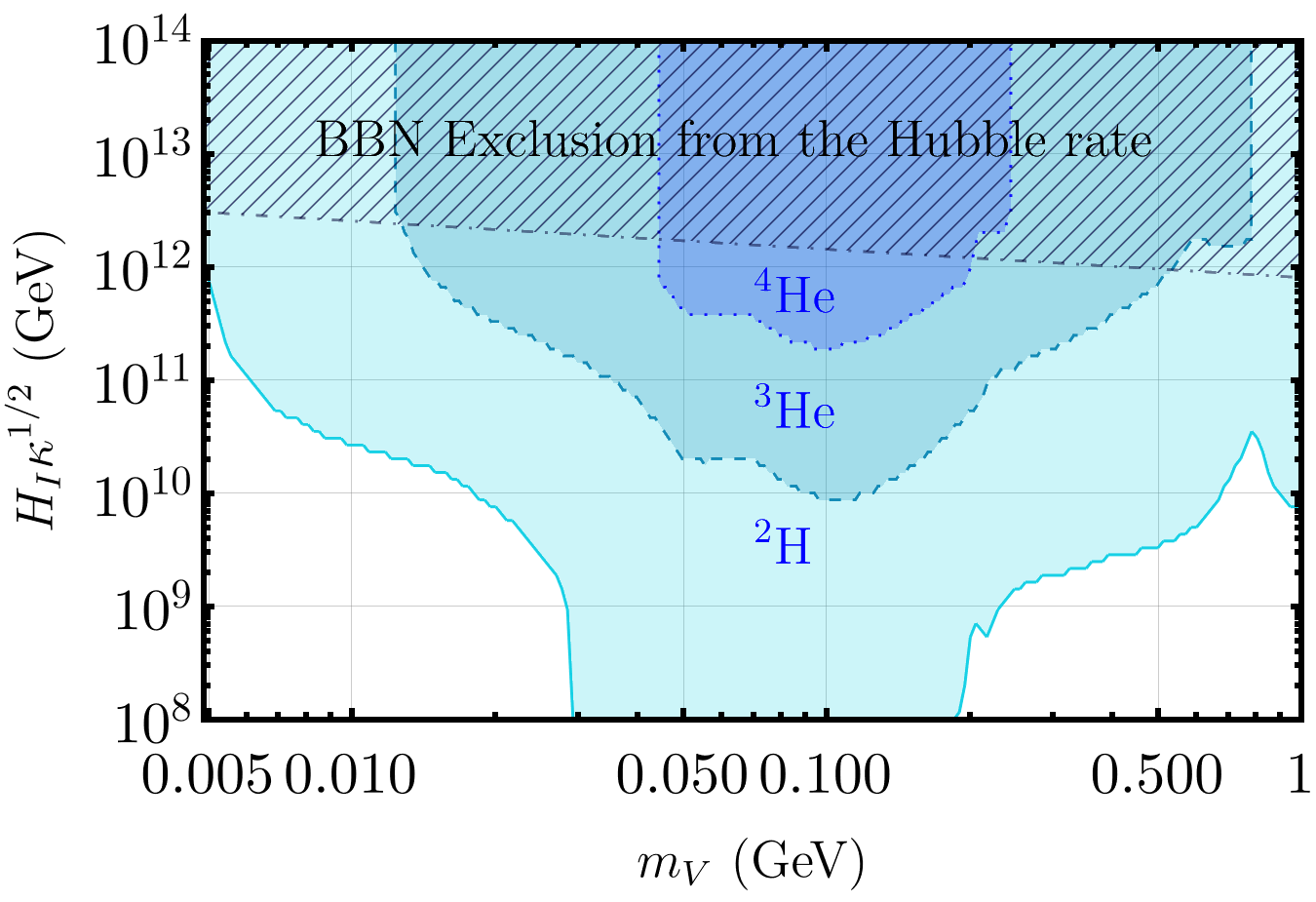}
    }
    \subfloat[\label{moneyplotb}]{
        \includegraphics[width=0.49\textwidth]{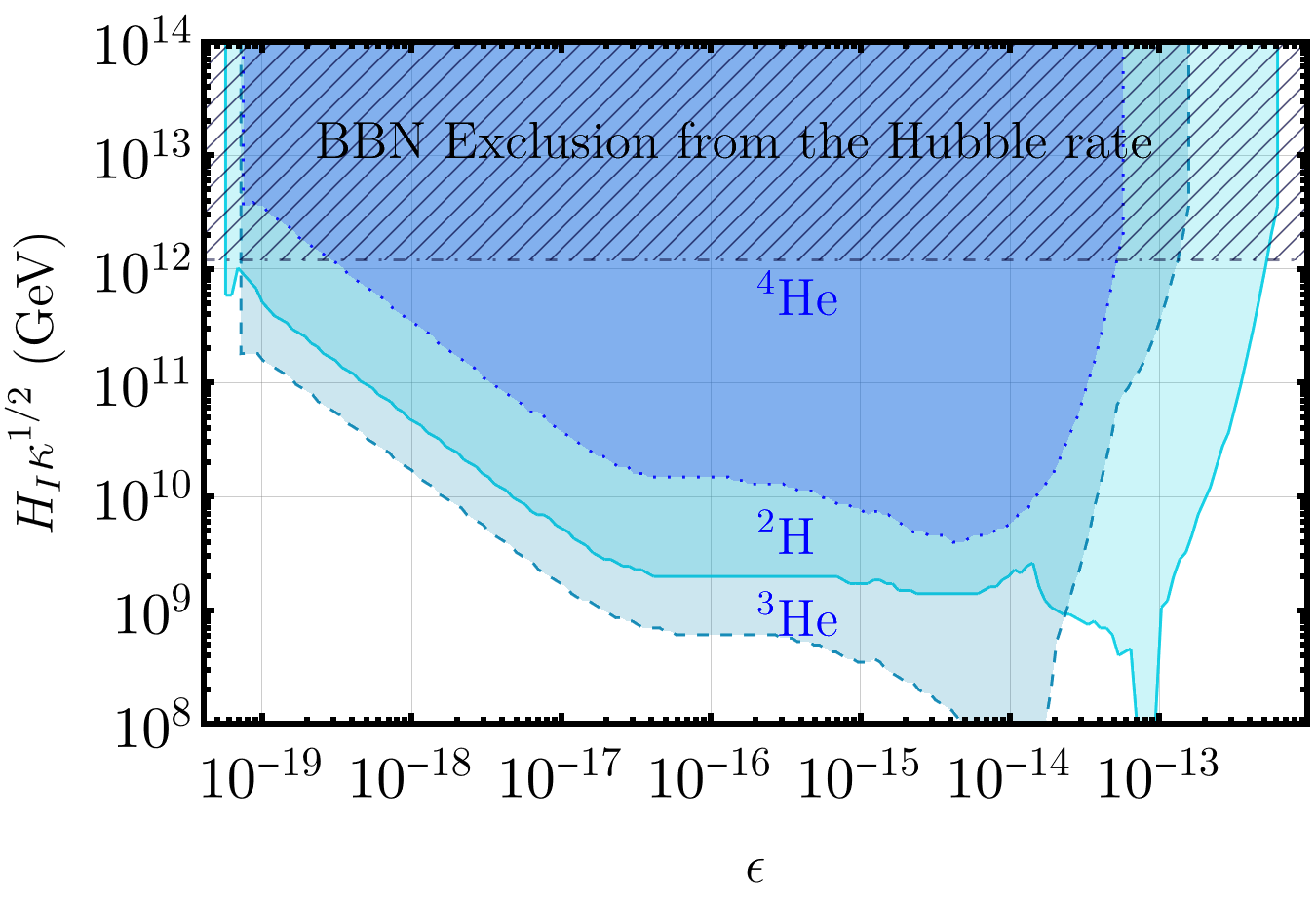}
    }
\caption{\emph{Left panel:} The BBN constraints on a dark photon with mass $m_V$ with kinetic mixing parameter to the SM photon $\epsilon = 5 \times 10^{-14}$ for high reheating scenario $T_{\rm RH} \lesssim 10^{15}$ GeV. $H_I$ is the Hubble scale at the end of inflation, while $\kappa$ is an order of one parameter to take into account the dependence on inflationary models. The dashed area ``BBN exclusion from the Hubble rate'' is ruled out at 95\% CL since the energy density of the dark photon is large enough to affect the BBN through the modification of the Hubble rate during the BBN. Further 95\% CL exclusion regions delimited by solid, dashed, and dotted lines are due to the BBN photodisintegration constraints from the measurements of primordial abundances of $^2$H, $^3$He, and $^4$He, respectively.  \textcolor{black}{\emph{Right panel:} Exclusion regions for $m_V = 0.2$ GeV in the plane of $H_I \kappa^{1/2}$ and $\epsilon$.}
}
\label{fig:ExclReg}
\end{figure}

\textcolor{black}{To illustrate the main idea of this work, we give a preview of our results in Figure \ref{fig:ExclReg}, which shows the BBN constraints on (a) $H_I \kappa^{1/2}$ vs the mass $m_V$ of dark photon plane, and (b) $H_I \kappa^{1/2}$ vs the kinetic mixing parameter $\epsilon$ plane. Here $\kappa$ is an order of one parameter to take into account the dependence on inflationary models. In Figure \ref{moneyplota}, we fix the kinetic mixing parameter of the dark photon with the SM photon to be $\epsilon = 5\times 10^{-14}$, and in Figure \ref{moneyplotb} we fix the mass of the dark photon at $m_V = 0.2$ GeV. In both cases we consider a high reheating scenario $T_{\rm RH} \lesssim 10^{15}$ GeV and account for both gravitational and freeze-in productions of the dark photon.} For large $H_I$ where gravitational production is dominant, the dashed area ``BBN exclusion from the Hubble rate'' is ruled out at 95\% CL since the energy density of dark photon is large enough to modify the Hubble rate during the BBN and thereby affect the BBN itself (see further discussion in Section \ref{subsec:cosmological_constraints}).
The different shaded areas delimited by solid, dashed and dotted lines are ruled out at 95\% CL due to the BBN photodisintegration constraints from the measurements of primordial abundances of $^2$H, $^3$He, and $^4$He, respectively (further details are in Section \ref{sec:constraints_dark_photon}). In summary, for high $H_I$, gravitational production of the dark photon is very efficient, ruling out entire light mass ranges of the dark photon. Eventually for $H_I \lesssim 10^8$ GeV, gravitational production becomes subdominant and we are left with the constraint from $^2$H coming mainly from the freeze-in production of the dark photon. Hence an eventual determination of $H_I$ from primordial gravitational wave contribution to the Cosmic Microwave Background (CMB) could greatly constrain the parameter space of new vector bosons.
\begin{comment}
\begin{figure}[th!] \label{fig3}
    \centering
    \includegraphics[width=0.65\textwidth]{MoneyPlot2.pdf}
\label{fig:ExclReg2}
\end{figure}
\end{comment}

This article is organized as follows. In Section \ref{sec:vector_model}, we discuss the a general vector boson $V$ that couples to the SM fermions and the possible decay channels of $V$ with $m_V \leq {\rm GeV}$. The details of photon and electron/positron spectra that result from the decays of $V$ are relegated to Appendix \ref{appendix-B}. 
In Section \ref{sec:gravi_prod}, we discuss gravitational production and the associated cosmological constraints. In Section \ref{sec:BBN_photo}, we derive the constraints from the photodisintegration effects on the light element abundance for a dark photon model. In Appendix \ref{appendix-A}, we discuss the model file that we have created (available on GitHub at \url{https://github.com/shengfong/lightvectorboson}) for a general light vector boson to be used with \texttt{ACROPOLIS}.  Finally in Section \ref{sec:conclusion_future}, we conclude and discuss some future directions.

\vspace{-1mm}
%%%%%%%%%%%%%%%%%%%%%%%%%%%%%%%%%%%%%%%%
%%%%%%%%%%%%%%%%%%%%%%%%%%%%%%%%%%%%%%%%
\section{General Light Vector Boson Model}\label{sec:vector_model}
In this work, we focus on new vector bosons in the mass range
$2m_e \leq m_V \leq 1\,{\rm GeV}$
that couple very feebly with the SM sector such that they never reach thermal equilibrium.
Even though we do not specify the origin of the vector boson mass, however, it can have a St\"{u}ckelberg mass~\cite{Stueckelberg:1938hvi}, or a mass generated by the Higgs mechanism~\cite{Higgs:1964pj} with a sufficiently heavy Higgs boson that fully decouples.  Its coupling with the SM fermions can be written as,    
\begin{align}
-\mathcal L &\supset 
\overline f \gamma^\mu  \left(g_V Q_{X,f} + e \epsilon Q_{\textrm{em},f} \right) P_X\; f\; V_\mu \;,
\end{align}
where, $X=L,R$ and $P_{L,R}= (1\mp \gamma_5)/2$. In the above equation, sum over all fermions $f$ must be understood. 
$Q_{\textrm{em},f}$ is the electric charge of the fermion $f$ in units of the proton charge $e$. Similarly, $Q_{X,f}$  is defined as the charge of the fermion $f_X$ under a new gauge symmetry $U(1)$.   For dark photon, there is no direct-coupling to SM fermions, hence $g_V=0$. However, it couples to the SM via kinetic mixing with the ordinary photon, which originates from the following Lagrangian: 
\begin{align} 
-\mathcal L \supset \frac{\epsilon}{2}F_{\mu \nu} V^{\mu \nu}\;,
\label{eq:kinetic_mixing}
\end{align}
where $\epsilon$ is the dimensionless kinetic mixing parameter, and $F_{\mu \nu} \equiv \del_\mu A_\nu - \del_\nu A_\mu$ ($V_{\mu \nu} \equiv \del_\mu V_\nu - \del_\nu V_\mu$) is the field strength of the SM (dark) photon.

On the other hand, if the vector boson is associated with a $U(1)$ symmetry under which SM fermions carry non-trivial charges $Q_{X,f}\neq 0$, then the dominant coupling arises from the $g_V$ term in the above equation. Examples of such $U(1)$'s are the anomaly-free combinations~\cite{He:1990pn,He:1990pn,He:1991qd} $B-L$~\cite{Heeck:2014zfa}, $L_{e}-L_{\mu/\tau}$~\cite{Wise:2018rnb}, and  $L_{\mu}-L_{\tau}$~\cite{Altmannshofer:2014pba}. In the case of $B-L$, right-handed neutrinos are required to be added to cancel gauge anomalies. By extending the SM fermion sector, a new baryonic force $U_B(1)$~\cite{Nelson:1989fx} can also be made anomaly free~\cite{Foot:1989ts}. In this work, any beyond the Standard Model (BSM) degrees of freedom, except the vector boson, are assumed to be heavy and decoupled from the low energy theory.   In general, these direct-coupling  vector bosons can  
also have a gauge kinetic mixing with the photon, which is assumed to be sub-leading compared to the direct gauge interaction. 
We are interested in sub-GeV dark photon (direct-coupling vector boson) for which $\epsilon\ll 1$ ($g_V\ll 1$) such that the rate at which these vector bosons interact with the SM particles is much less than the Hubble rate and never reach thermal equilibrium with the SM bath.
However, vector bosons of these types can still be abundantly produced via gravitational interactions, as discussed in Section \ref{sec:gravi_prod}. Moreover, in part of the parameter space, freeze-in production may also be relevant, which we add to the purely gravitational production to obtain the total abundance.  

In the following two subsections, we review the decay modes of the vector bosons and the primary photon and electron/positron spectra they produce, respectively, essential in computing the associated BBN photodisintegration bounds.

%%%%%%%%%%%%%%%%%%%%%%%%%%%%%%%%%%%%%%%%
\subsection{Decay widths} \label{decaywidths}
We assume that the vector bosons decay into SM particles and ignore any decay into exotic BSM particles.   Under the above assumptions, the sub-GeV vector bosons can have two leptonic decay modes of interest, and three hadronic decay modes, as described below.

%%%%%%%%%%%%%%%%%%%%%%%%%%%%%%%%%%%%%%%%
\subsubsection{Leptonic decay}
For $m_V\leq\,\textrm{GeV}$, only two leptonic decay channels $V \rightarrow e^+ e^-$ and $V \rightarrow \mu^+ \mu^-$ are possible, and their decay widths are given by 
\begin{align}\label{decayw}
    \Gamma (V \rightarrow \bar{\ell}\ell) = \frac{1}{6} \alpha_V m_V \sqrt{1-\frac{4m_\ell^2}{m_V^2}} \left[ Q_{L,\ell}^2 + Q_{R,\ell}^2 - \frac{m_\ell^2}{m_V^2} \left( Q_{L,\ell}^2 - 6Q_{L,\ell} Q_{R,\ell} + Q_{R,\ell}^2 \right) \right], 
\end{align}
where  $\alpha_V = \epsilon^2 \alpha \simeq \epsilon^2/137$ 
and $Q_{L,R} = Q_{\textrm{em}}$ for dark photons, and 
$\alpha_V = g_V^2/(4\pi)$
for direct-coupling vector bosons.
Since $Q_{\textrm{em},\nu}=0$, 
decay to neutrinos does not happen for the dark photon. However, decays to neutrinos are present for direct-coupling vectors and can have large branching fractions. For $B-L$ theory, in the entire mass range, the branching ratio to neutrinos dominates over the rest, except near the $\omega$-resonance, which turns on at $m_{\omega}=782$ MeV that opens up hadronic decays~\cite{Bauer:2018onh}. As mentioned above, vector boson decays to BSM particles (such as right-handed neutrinos in the case of  $B-L$) are kinematically forbidden. For $L_e-L_{\mu/\tau}$ theories, the branching ratio to neutrinos can be as large as to charged leptons; however, for $L_\mu-L_\tau$, below muon threshold, decay is fully dominated by $V\to\nu\overline \nu$ by several orders of magnitude~\cite{Bauer:2018onh}. Even though leptons do not carry any charge under the $U_B(1)$, the baryonic vector boson is not completely decoupled from the leptons due to its possible  kinetic mixing ($\epsilon$) with the photon.  Even if it is set to zero at the tree-level, one-loop radiative corrections involving heavy quarks lead to $\epsilon \neq 0$~\cite{Tulin:2014tya,Carone:1995pu,Aranda:1998fr}. \textcolor{black}{The size of this induced coupling is given by the product of the relevant gauge couplings and suppressed by the loop factor, $\epsilon\sim e g_V/(16\pi^2)\sim 6.7\times 10^{-3} \alpha^{1/2}_V$. } With this, its decay width to leptons is identical to the dark photon scenario.

\vspace{0.2cm}
%%%%%%%%%%%%%%%%%%%%%%%%%%%%%%%%%%%%%%%%
\subsubsection{Hadronic decay}
Since hadronic decays of $L_\ell - L_{\ell^\prime}$ vector bosons are always suppressed, we only consider the decays of the dark photon, $B-L$, and baryonic vector bosons. 
Their decays into specific hadronic states occur via induced kinetic mixing with the $\rho$ and $\omega$ vector mesons as well as via direct coupling to the electromagnetic (baryonic) currents in the case of dark photons ($B-L$ vector bosons). Using the Vector Meson Dominance (VMD) model~\cite{Sakurai:1960ju, Fujiwara:1984mp, Bando:1984ej, OConnell:1995nse, Tulin:2014tya}, the $\rho$- and $\omega$-mesons are treated as gauge bosons of a hidden $U(2)$ flavor symmetry. The approximate strength of the induced kinetic
mixing with $\rho$ and $\omega$ vector mesons are determined by
\begin{align}
    2 \operatorname{tr}\left(t_{A} Q_{V}\right) \sqrt{4\pi \alpha_V},
\end{align}
where the $U(2)$ generators are $t_A = \text{diag}(1/2, \mp 1/2)$ for $A = \rho, \omega$, respectively, and  $Q_V = \textrm{diag}(Q_{V,u},Q_{V,d})$, which for the dark photon corresponds to $Q_{V,q}\to Q_{\textrm{em},q}$.

\vspace{0.2cm}
%%%%%%%%%%%%%%%%%%%%%%%%%%%%%%%%%%%%%%%%
\noindent {$\pmb{V \rightarrow \pi^+ \pi^-}$}\\
The leading hadronic decay $V \rightarrow \pi^+ \pi^-$ of the dark photon  arises from the direct contribution of the charged pions to the electromagnetic current as well as from the induced kinetic mixing with the $\rho$-meson. The dominant contribution of the baryonic vector boson arises from its direct mixing with the $\omega$-meson.  This decay width can be written as~\cite{Tulin:2014tya},
\begin{align}
    \Gamma\left(V \rightarrow \pi^{+} \pi^{-}\right)=\frac{\epsilon^2\alpha}{12}m_{V}\left(1-\frac{4 m_{\pi^{\pm}}^2}{m_{V}^2}\right)^{3 / 2}\left|F_{\pi \pi}\left(m_{V}^2\right)\right|^{2},
\end{align}
where $m_{\pi^\pm} = 139.57$ MeV.
The form factor can be expressed as~\cite{Tulin:2014tya}
\begin{align}
    F_{\pi \pi} (s) = F_\rho(s) \left( 1+ \frac{1+\delta}{3} \frac{\tilde{\Pi}_{\rho \omega}(s)}{s-m_\omega^2+i m_\omega \Gamma_\omega}  \right),\label{FORMpipi}
\end{align}
where $F_\rho$ is the form factor due to $\rho$ exchange only, and can be approximated with a simple Breit-Wigner form~\cite{Tulin:2014tya}
\begin{align}
    F_\rho(s) \approx \left(1-\frac{s}{m_{\rho}^2}- i \frac{\Gamma_\rho}{m_\rho}\right)^{-1},
\end{align}
with $m_\omega = 782.65$ MeV, $m_\rho = 775.25$ MeV, $\Gamma_\omega = 8.49$ MeV, and $\Gamma_\rho = 149$ MeV. $\tilde{\Pi}_{\rho \omega}$ is the additional isospin-violating $\rho-\omega$ mixing term, which takes the value $\tilde{\Pi}_{\rho \omega}(s) \simeq \tilde{\Pi}_{\rho \omega}(m_\omega^2) = -3500 \pm 300\ \text{MeV}^2$~\cite{Gardner:1997ie}. Since there is no direct coupling, in Eq.~\eqref{FORMpipi}, $\delta$ vanishes for the case of dark photon. On the other hand, this quantity for the baryonic vector boson is large since $\delta=2g_V/(e\epsilon)\approx 4\pi/\alpha\gg 1$~\cite{Tulin:2014tya}.

\vspace{0.2cm}
%%%%%%%%%%%%%%%%%%%%%%%%%%%%%%%%%%%%%%%%
\noindent $\pmb{V \rightarrow \pi^0 \gamma}$\\
The decay into $\pi^0 \gamma$ is facilitated by the dark photon, $B-L$ and baryonic vector bosons mixing with the $\omega$-meson.  The decay width for this channel is given by~\cite{Tulin:2014tya}
\begin{align}
    \Gamma(V \rightarrow \pi^0 \gamma) = \left[2\text{tr}(t_A Q_V) \right]^2 \frac{3\alpha\; \alpha_V}{128 \pi^3} \frac{m_V^3}{f_\pi^2} \left|F_\omega(m_V^2)\right|^2,
\end{align}
where $f_\pi = 93$ MeV is the pion decay constant, and the form factor is given by~\cite{Tulin:2014tya}
\begin{align}
    F_\omega(s) \approx \left( 1- \frac{s}{m_\omega^2}-i\frac{\Gamma_\omega}{m_\omega} \right)^{-1}.
\end{align}

\vspace{0.2cm}
%%%%%%%%%%%%%%%%%%%%%%%%%%%%%%%%%%%%%%%%
\noindent $\pmb{V \rightarrow \pi^+ \pi^- \pi^0}$\\
This channel is also facilitated by  mixing with the $\omega$-meson, which decays into pions via coupling with the $\rho$-meson.
The decay width can be expressed as~\cite{Tulin:2014tya}
\begin{align}
    \Gamma(V\rightarrow \pi^+ \pi^- \pi^0) = \left[2\text{tr}(t_A Q_V) \right]^2 \frac{3 \alpha_V}{16 \pi^4} \left( \frac{g_{\rho \pi \pi}^2}{4\pi} \right)^2 \frac{m_V}{f_\pi^2} \mc I(m_V^2) \left|F_\omega(m_V^2)\right|^2,\label{decaywidth3pion}
\end{align}
where the $\rho\pi\pi$ coupling is fixed by $g_{\rho \pi \pi}^2/(4\pi) \simeq 3$ to reproduce the observed $\rho \rightarrow \pi \pi$ decay rate. The integral over the phase space is given by~\cite{Tulin:2014tya}
\begin{align}
    \mc I(m_V^2) = \int dE_+ dE_- &\left[ |\mathbf{p}_+|^2 |\mathbf{p}_-|^2 - (\mathbf{p}_+ \cdot \mathbf{p}_-)^2 \right] \nonumber \\
    &\times \left( \frac{1}{m_\rho^2 - (p_+ + p_-)^2} + \frac{1}{m_\rho^2 - (p_0 + p_+)^2} + \frac{1}{m_\rho^2 - (p_0 + p_-)^2} \right)^2,\label{integral3pion}
\end{align}
where the pion momenta $4$-vectors in the vector boson's rest frame ($p_V \equiv (E_V, \mathbf{0})$) are defined as $p_i \equiv (E_i, \mathbf{p}_i)$. All quantities in the integrand are expressed in terms of $E_\pm$, $m_V$ and the pion masses $m_{\pi^\pm} = 139.57$ MeV, $m_{\pi^0} = 134.98$ MeV,
\begin{align}
    \mathbf{p}_+ \cdot \mathbf{p}_- &= m_{\pi^\pm}^2 + E_+ E_- - m_V (E_+ + E_-) + \frac{1}{2}(m_V^2 - m_{\pi^0}^2), \\
    (p_+ + p_-)^2 &= (p_V - p_0)^2 = -m_V^2 + m_{\pi^0}^2 + 2m_V (E_+ + E_-), \\
    (p_0 + p_+)^2 &= (p_V - p_-)^2 = m_V^2 + m_{\pi^-}^2 - 2m_V E_-, \\
    (p_0 + p_-)^2 &= (p_V - p_+)^2 = m_V^2 + m_{\pi^+}^2 - 2m_V E_+.
\end{align}
The phase space is limited to the kinematically allowed domain, which can be expressed as $\int dE_+ dE_- \equiv \int_{m_{\pi^\pm}}^{\mc E_*} dE_+ \int_{\mc E_-}^{\mc E_+} dE_- $, where
\begin{align}
    \mc E_* = \frac{1}{2} \left( m_V - 3\frac{m_{\pi^\pm}^2}{m_V}\right), \quad \mc E_{\pm} = \frac{1}{2} \left( m_V - E_+ \pm |\mathbf{p}_+| \sqrt{\frac{m_V^2 - 2E_+ m_V - 3m_{\pi^\pm}^2}{m_V^2 - 2E_+ m_V + m_{\pi^\pm}^2}} \right),
\end{align}
and in the relations above, we have approximated $m_{\pi^0} \simeq m_{\pi^{\pm}}$.

\textcolor{black}{Branching ratios into these five decay channels discussed in this section are model dependent. An example for the dark photon model will be discussed in Section \ref{sec:dark_photon}.}

%%%%%%%%%%%%%%%%%%%%%%%%%%%%%%%%%%%%%%%%
%%%%%%%%%%%%%%%%%%%%%%%%%%%%%%%%%%%%%%%%
\subsection{Electromagnetic spectra}
For a sub-GeV vector boson under consideration, it decays to leptons $V\to \ell \bar{\ell}$ as well as to the lightest hadrons $V\to \pi^0\gamma, \pi^+\pi^-, \pi^+\pi^-\pi^0$. The final daughter particles resulting  from these decay modes are the photons,  electrons, and positrons (as well as neutrinos). The full energy spectra of photons and electrons per decay are then obtained by summing over all decay modes.  All necessary details required to compute the electromagnetic spectra from the above-mentioned decay channels are summarized in Appendix~\ref{appendix-B}. We apply these injection spectra to compute the full electromagnetic cascade spectrum produced by the vector boson decay for the BBN photodisintegration studies.

%%%%%%%%%%%%%%%%%%%%%%%%%%%%%%%%%%%%%%%%
%%%%%%%%%%%%%%%%%%%%%%%%%%%%%%%%%%%%%%%%
\section{Gravitational Production of Massive Vector Bosons}\label{sec:gravi_prod}
In this section, we will first review the reheating of the Universe, the gravitational production of massive vector boson based on Ref.~\cite{Graham:2015rva,Ema:2019yrd,Ahmed:2020fhc,Kolb:2020fwh} (we closely follow Ref.~\cite{Kolb:2020fwh}) and then discuss the relevant cosmological constraints. 

%%%%%%%%%%%%%%%%%%%%%%%%%%%%%%%%%%%%%%%%
\subsection{Reheating}
Assuming immediate reheating after inflation, the energy density of the inflaton field $\rho_I = 3H_{I}^{2}M_{{\rm Pl}}^{2}$ is converted to the radiation density $\rho_R = \frac{\pi^{2}}{30}g_{\star}T^{4}$ at temperature $T_{\rm RH}$
%%%
\begin{eqnarray}
3H_{I}^{2}M_{{\rm Pl}}^{2} & = & \frac{\pi^{2}}{30}g_{\star{\rm RH}}T_{{\rm RH}}^{4},
\end{eqnarray}
%%%
where $M_{{\rm Pl}}=2.4\times10^{18}$ GeV is the reduced Planck mass
and $g_{\star{\rm RH}}\equiv g_{\star}\left(T_{{\rm RH}}\right)=106.75$
for the SM relativistic degrees of freedom. Solving for $T_{\rm RH}$, it will be the maximum reheating temperature that can be achieved
%%%
\begin{eqnarray}
T_{{\rm RH}}^{{\rm max}} & = & 8.4\times10^{14}\,{\rm GeV}\left(\frac{106.75}{g_{\star{\rm RH}}}\right)^{1/4}\left(\frac{H_{I}}{10^{12}\,{\rm GeV}}\right)^{1/2}.
\label{eq:TRH_max}
\end{eqnarray}
%%%
In the single-field model of inflation, the CMB limit on gravitational wave contribution gives $H_{I}\lesssim3\times10^{14}$ GeV. 

If reheating is not instantaneous, the final reheating temperature
when $\left.\rho_{I}\right|_{T_{\rm RH}}=\rho_{R}$ can be much smaller than $T_{{\rm RH}}^{{\rm max}}$.
Assuming matter domination during reheating $H\propto a^{-3/2}$ where $a$ is the cosmic scale factor, we have 
%$H_{{\rm RH}}/H_{I}=\left(a_{I}/a_{{\rm RH}}\right)^{3/2}$ and hence
%%%
\begin{equation}
    H_{\rm RH}^2 = H_I^2 \left(\frac{a_I}{a_{\rm RH}}\right)^3,
    \label{eq:reheating}
\end{equation}
and hence
\begin{eqnarray}
T_{{\rm RH}} & = & \frac{8.4\times10^{14}\,{\rm GeV}}{\alpha_{\rm RH}^{3/4}}\left(\frac{106.75}{g_{\star{\rm RH}}}\right)^{1/4}\left(\frac{H_{I}}{10^{12}\,{\rm GeV}}\right)^{1/2}, \label{eq:TRH}
\end{eqnarray}
%%%
where we have defined $\alpha_{{\rm RH}}\equiv a_{{\rm RH}}/a_{I} \geq 1$.

After reheating, the Universe is radiation dominated and the cosmic entropy density is given by
%%%
\begin{eqnarray}
s & = & \frac{2\pi^{2}}{45}g_{\star s}T^{3}=\frac{4\rho_{R}}{3T}\frac{g_{\star s}}{g_{\star}}. \label{entropy}
\end{eqnarray}
From the conservation of entropy $s a^3$ after reheating, we have
%%%
\begin{equation}
sa^{3} = s_{{\rm RH}}a_{{\rm RH}}^{3}=\left.\frac{4\rho_{R}}{3T}\frac{g_{\star s}}{g_{\star}}\right|_{T=T_{{\rm RH}}}a_{{\rm RH}}^{3}
=  \frac{4M_{{\rm Pl}}^{2}H_{{\rm RH}}^{2}}{T_{{\rm RH}}}a_{{\rm RH}}^{3}
% = \frac{4M_{{\rm Pl}}^{2}}{T_{{\rm RH}}}H_{I}^{2}a_{I}^{3}
 = \frac{4M_{{\rm Pl}}^{2}}{T_{{\rm RH}}H_{I}},
\end{equation}
where in the second equality, we have taken $\frac{g_{\star s}}{g_{\star}}\left|\right._{T=T_{{\rm RH}}}=1$
and in the last equality, we have used Eq.~\eqref{eq:reheating} and set the cosmic scale at the end of inflation to be $H_{I}a_{I}=1$. 

In the following, we will not consider instantaneous reheating and treat both $H_I$ and $T_{\rm RH}$ as free parameters with $T_{\rm RH}$ (equivalently $\alpha_{\rm RH}$) subject to the upper bound Eq.~\eqref{eq:TRH_max}. As shown in Ref.~\cite{Giudice:2000ex}, during the period of reheating, the thermal bath can achieve a maximum temperature 
%%%
\begin{equation}
T_{\rm max} = \left(\frac{3}{8}\right)^{2/5}  \left(\frac{40}{\pi^2}\right)^{1/8}
\frac{g_{\star{\rm RH}}^{1/8}}{g^{1/4}_{\star}(T_{\rm max})}
\left(M_{\rm Pl} H_I T_{\rm RH}^2\right)^{1/4},
\end{equation}
%%%
where $T_{\rm max} \leq T_{\rm RH}^{\rm max}$.
In the following, we will assume $g_{\star}(T_{\rm max}) = g_{\star{\rm RH}}$ and we can rewrite
%%%
\begin{equation}
T_{\rm RH} = \frac{3.1 \times 10^{14}\,{\rm GeV}}{r_T^2}
\left(\frac{106.75}{g_{\star \rm RH}}\right)^{1/4}\left(\frac{H_I}{10^{12}\,{\rm GeV}}\right)^{1/2}, \label{eq:TRH2}
\end{equation}
%%%
where we have defined the ratio $r_T \equiv T_{\rm max}/T_{\rm RH} \geq 1$. Comparing the equation above taking $r_T = 1$ with Eq.~\eqref{eq:TRH}, we see that the temperature of immediate reheating $\alpha_{\rm RH} = 1$ is not achievable once the finite period of reheating is taken into account. 
\textcolor{black}{Given a $H_I$, $T_{\rm RH}$ can be much smaller than $T_{\rm max}$ if the decay rate of the inflaton, $\Gamma_\phi$, into radiation  is small since at the reheating $H(T_{\rm RH}) = \Gamma_\phi$.}
For example, taking $H_I = 10^8\,{\rm GeV}$, $r_T \lesssim 3\times 10^7$ such that $T_{\rm RH} \gtrsim 4\,{\rm MeV}$ from the BBN bound~\cite{deSalas:2015glj}.

%%%%%%%%%%%%%%%%%%%%%%%%%%%%%%%%%%%%%%%%
\subsection{Gravitational production}

For a massive vector field, there are three degrees of freedom: two transverse components and one longitudinal component. The transverse components, in the massless limit, couple conformally to the gravity and will not be produced. The longitudinal component, on the other hand, will be produced even in the massless limit since it corresponds to a minimally coupled scalar field. For $m_V < H_I$ where $H_I$ is the Hubble rate at the end of inflation, the production of the longitudinal component will be the dominant one and will be considered here~\cite{Graham:2015rva,Ema:2019yrd,Ahmed:2020fhc,Kolb:2020fwh}. 

We can express the final abundance of vector boson $V$ produced as
%%%
\begin{eqnarray}
Y_V & = & \frac{na^{3}}{sa^{3}} = \frac{T_{{\rm RH}}H_{I}}{4M_{{\rm Pl}}^{2}}\int\frac{dk}{k}n_{k},
\end{eqnarray}
%%%
where $n_k$ is the spectrum of the mode function of the two transverse components and a longitudinal component of a massive vector boson field. 
\textcolor{black}{As first pointed out in Ref.~\cite{Graham:2015rva}, unlike a minimally coupled scalar where $n_k$ is constant for long wavelength mode i.e. small $k$, the longitudinal component of vector field is suppressed when the modes are nonrelativistic and this leads to $n_k \propto k^2$ for small $k$. If $V$ were to be dark matter, this is crucial to suppress large-scale isocurvature perturtations, allowing it to constitute a good dark matter candidate~\cite{Graham:2015rva}.}   In Ref.~\cite{Kolb:2020fwh}, considering only the longitudinal component and assuming that the Universe is de-Sitter during inflation, matter-dominated during reheating and radiation-dominated after reheating, an approximate analytic solution for a vector boson with mass $m_V < H_I$ from gravitational production is given by\footnote{\textcolor{black}{Ref. \cite{Ahmed:2020fhc} showed that an equation of state different from matter domination during reheating $w \neq 0$ will affect the short wavelength modes. However, for light vector bosons considered here, the effect is suppressed by $\sqrt{m_V/H(T_{\rm RH})}$, and we will leave this exploration for future study when we consider heavier vector bosons.}}
%%%
\begin{eqnarray} \label{eq:gravyield}
Y_{V} & = & \kappa\frac{H_I^2}{16\pi^2 M_{\rm Pl}^2}\begin{cases}
\dfrac{b}{\sqrt{m_{V}}}\left(\dfrac{3}{2}-\dfrac{2}{3}\dfrac{b\sqrt{m_{V}}}{T_{{\rm RH}}}-\dfrac{1}{3}\dfrac{T_{{\rm RH}}\sqrt{m_{V}}}{bH_{I}}\right) & T_{{\rm RH}}>b\sqrt{m_{V}} \\
\dfrac{T_{{\rm RH}}}{m_{V}}\dfrac{5}{6}\left(1-\dfrac{2}{5}\dfrac{m_{V}}{H_{I}}\right) & T_{{\rm RH}}<b\sqrt{m_{V}}
\end{cases},
\end{eqnarray}
%%%
where for the two cases above, reheating completes after (before) the longitudinal mode becomes nonrelativistic. We have defined $b\equiv\left(\frac{\pi^{2}}{90M_{{\rm Pl}}^{2}}g_{\star{\rm RH}}\right)^{-1/4}$, and $\kappa \sim 1-10$ is an order of one factor which captures the dependence on the inflationary models \cite{Kolb:2020fwh}. 
For a given $H_I$, one has the freedom to choose $T_{\rm RH}$ according to Eq.~\eqref{eq:TRH2} with $r_T \geq 1$. In this work, we will always take $g_{\star\rm RH} = 106.75$. In our study, we will consider two illustrative scenarios, a high reheating scenario with $r_T = 1$ and a low reheating scenario with $r_T = 10^6$:
%%%
\begin{eqnarray} \label{eq:Ygrav_2cases}
m_V Y_{V} & \simeq & \kappa \begin{cases}
1.4 \times 10^{-7}\,{\rm GeV} 
\left(\dfrac{H_I}{10^{12}\,{\rm GeV}}\right)^{2} 
\left(\dfrac{m_V}{10\,{\rm MeV}}\right)^{1/2}, 
& r_T = 1 \\
2.8 \times 10^{-8}\,{\rm GeV} \left(\dfrac{H_I}{10^{14}\,{\rm GeV}}\right)^{5/2}, & r_T = 10^6
\end{cases},
\end{eqnarray}
%%%
where the approximate expressions are valid for $0.1\,{\rm eV} \lesssim m_V \lesssim 10^6\,{\rm GeV}$ and $10^7\,{\rm GeV} \lesssim H_I \lesssim 10^{15}\,{\rm GeV}$. 
In this work, we will focus on light vector boson $2m_e \leq m_V \leq 1\,{\rm GeV}$ and large $H_I \gtrsim 10^8\,{\rm GeV}$.

%%%%%%%%%%%%%%%%%%%%%%%%%%%%%%%%%%%%%%%%%%%%%
\subsection{Cosmological constraints}
\label{subsec:cosmological_constraints}

Assuming the new vector boson $V$ to couple only to the SM fields as discussed in the previous section, if $m_{V} \leq 2 m_e$  and/or if its lifetime is greater than the age of the Universe $\tau > 4.4\times 10^{17}\,{\rm s}$, $V$ can constitute a good dark matter due to the suppression of isocurvature perturbations~\cite{Graham:2015rva}. In that case, one should impose an upper bound on its energy density 
%%%
\begin{eqnarray}
\Omega_{V}h^2 = \frac{m_{V}Y_{V} s_0 h^2}{\rho_c} \leq \Omega_{\rm DM}h^2
\end{eqnarray}
%%%
where $\Omega_{\rm DM}h^2 = 0.12$ is the observed dark matter abundance from Planck~\cite{Planck:2018vyg} with $\rho_c \simeq 1.05 \times 10^{-5} h^2$~GeV/cm$^3$ the present cosmic critical energy density, and $s_0 \simeq 2.89 \times 10^3$~cm$^{-3}$ the present entropy density. The above translates to
%%%
\begin{equation}
    m_V Y_V \leq 4.36 \times 10^{-10}\,{\rm GeV}.
    \label{eq:DM_bound}
\end{equation}
%%%

\begin{figure}[!ht] 
    \centering
    \includegraphics[width=0.60\textwidth]{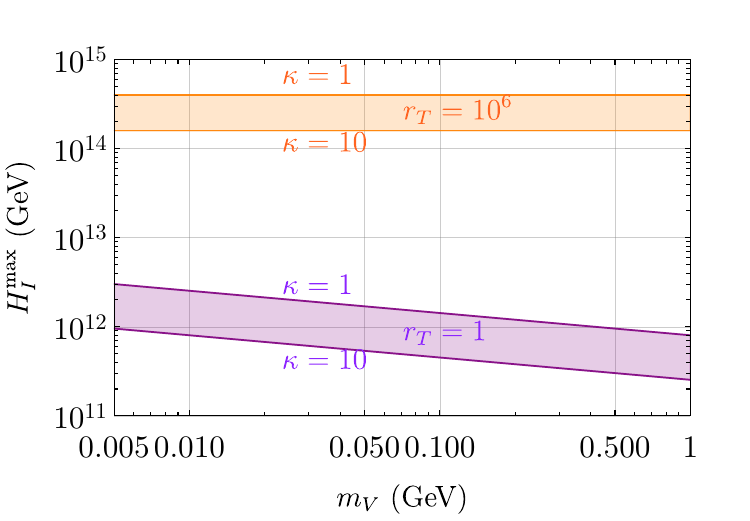}
    \caption{The upper bound on the Hubble rate after inflation $H_I^{\rm max}$ as a function of new vector boson mass $m_V$. The upper (lower) region corresponds to $r_T = 10^6$ ($r_T = 1$). In either case, the upper (lower) boundary of the regions correspond to $\kappa = 1$ ($\kappa = 10$).} \label{fig:myHImax}
\end{figure}

For $m_{V} > 2 m_e$, the bound will depend on when the $V$ particles decay. If all of them decay before the BBN, besides the possible entropy production (one has to take this into account such that the baryon and dark matter energy densities agree with observations), there is no other constraint. On the other hand, if $V$ particles decay in the epoch between the BBN and the CMB, their presence will change the Hubble rate throughout the BBN.
Using the public code \texttt{AlterBBN}~\cite{Arbey:2011nf,Arbey:2018zfh}, we included $V$ contribution to the Hubble rate throughout the BBN and found a constraint of 
%%%
\begin{equation}
m_V Y_V < 0.9 \times 10^{-6}\,{\rm GeV},
\label{eq:BBN_constraint}
\end{equation}
%%%
in order to be consistent with the deuterium and Helium-4 primordial abundances at 95\% CL. For a given $r_T$ and $m_V$, this implies an upper bound on $H_I$. In Figure \ref{fig:myHImax}, we show the upper bound on $H_I$ as a function of $m_V$ for $r_T = 1$ and $r_T = 10^6$. As we can see from Eq.~\eqref{eq:Ygrav_2cases}, $m_V Y_V$ is independent of $m_V$ for low reheating scenario $r_T = 10^6$. In the bands, $\kappa$ varies from 1 to 10 to capture the dependence on inflationary models. From Eqs.~\eqref{eq:Ygrav_2cases} and \eqref{eq:DM_bound}, if $V$ were to be a metastable dark matter in the mass range $2m_e \leq  m_V \leq 1$\,GeV, one requires,
%%%
\begin{alignat}{2}
1.8 \times 10^{10}\,{\rm GeV} \leq &H_I \kappa^{1/2} \leq 9.9\times 10^{10}\,{\rm GeV}, \;\;\;&&\textrm{for $r_T = 1$},
\label{eq:viable_DM}
\\&
H_I \kappa^{2/5} \leq 1.9\times 10^{13} \,{\rm GeV}, \;\;\;&&\textrm{for $r_T = 10^6$}.
\end{alignat}

Furthermore, in view of the consistency between the baryon density determined from the BBN and the Cosmic Microwave Background (CMB), we can impose a limit on the entropy production from the decays of $V$ at $T_d$. From energy conservation and assume instantaneous decays of $V$ at $T_d$, we have
%%%
\begin{equation}
    \frac{\pi^2}{30}g_\star T_d^4 + m_{V} Y_{V} s (T_d) 
    = \frac{\pi^2}{30}g_\star \tilde T^4 = \rho_r(\tilde T),
\end{equation}
%%%
where $\tilde T$ is the radiation temperature immediately after decay of $V$. The ratio of entropy before and after decay of $V$ will be
%%%
\begin{equation}
    \frac{S(\tilde T)}{S(T_d)} = \left(\frac{\tilde T}{T_d}\right)^3 
    = \left(1 + \frac{4}{3} \frac{m_{V}Y_{V}}{T_d}\right)^{3/4}.
\end{equation}
%%%
Assuming we can tolerate an entropy production of $100 \eta$ \% with $\eta \ll 1$, we obtain  
%%%
\begin{equation}
   T_d \gtrsim \frac{1}{\eta} m_{V} Y_{V}. \label{eq:decay_bound}
\end{equation}
%%%
The constraint above only applies when $\tilde T \gtrsim 0.26$ eV, i.e., the decay completes before the recombination or the formation of the CMB. This can be translated to the constraint on the decay rate of $V$ where we estimate $T_d$ from the condition
%%%
\begin{equation}
   \Gamma_{V} =  H(T_d) 
   = \sqrt{\frac{\rho_r(\tilde T)}{3M_{\rm Pl}^2}},
   \label{eq:decay_condition}
\end{equation}
%%%
which also gives the thermal temperature after decay
%%%
\begin{equation}
   \tilde T = \sqrt{\frac{90\Gamma_V M_{\rm Pl}}{\pi g_\star^2}}. % \gtrsim 0.26\,{\rm eV}.
   \label{eq:T_after_decay}
\end{equation}
%%%

Later, for our study on the photodisintegration effects of the BBN that takes place only after the BBN completes, we will consider only the scenarios with $H_I$ smaller than the upper bounds shown in Figure \ref{fig:myHImax} such that the modification to the Hubble rate due to the presence of $V$ is consistent with the BBN. Taking $V$ as a dark photon as an illustrative example in Section \ref{sec:dark_photon}, we will impose Eq.~\eqref{eq:decay_bound} with $\eta = 0.01$ to show the parameter space where the entropy dilution of the baryon energy density between the epoch of the BBN and the CMB is not more than 1\%.

%%%%%%%%%%%%%%%%%%%%%%%%%%%%%%%%%%%%%%%%
%%%%%%%%%%%%%%%%%%%%%%%%%%%%%%%%%%%%%%%%
\section{BBN Photodisintegration Constraints}
\label{sec:BBN_photo}

The BBN is one of the most important probes of BSM physics~\cite{Sarkar:1995dd,Iocco:2008va,Jedamzik:2009uy,Pospelov:2010hj}. In the standard cosmological scenario, from the thermal bath, neutrons  freeze out at a temperature $\sim 1$ MeV that corresponds to $t\sim 1$ s after the Big Bang,  which sets the initial condition for the
BBN. Then the light elements such as $^2\textrm{H}$, $^3\textrm{He}$, $^4\textrm{He}$, $^7\textrm{Li}$ are subsequently synthesized~\cite{Schramm:1977ne,Bernstein:1988ad,Walker:1991ap} and theoretical estimations of their primordial abundances are possible at a time  $t\sim 180$ s. When the standard BBN ends at  $t\sim 10^4$ s, the abundances of light elements are no longer changed by the nuclear fusion and spallation reactions.

Remarkably, the predictions of the standard cosmology are in excellent agreement~\cite{Cyburt:2015mya} with observations after taking into account the alteration due to the stellar nucleosynthesis when some light elements are ejected and some heavy elements are formed, which introduces some uncertainty. The presence of any exotic particle arising from the BSM model may alter the predicted abundances of these light elements. After the BBN ends, i.e., for $t\gtrsim 10^4$ s, further modifications of the nuclear abundances due to photodisintegration reactions are possible.  Models consisting of long-lived particles  are therefore highly constrained from the BBN measurements due to the energy they inject into the cosmological plasma~\cite{Ellis:1984er,Juszkiewicz:1985gg,Dimopoulos:1987fz,Reno:1987qw,Dimopoulos:1988ue,Frieman:1989fx,Ellis:1990nb,Moroi:1993mb,Kawasaki:1994af,Cyburt:2002uv,Jedamzik:2004er,Kawasaki:2004qu,Jedamzik:2006xz,Kawasaki:2008qe,Hisano:2008ti,Hisano:2009rc,Kawasaki:2015yya,Cyburt:2004yc,Ho:2012ug,Boehm:2013jpa,Nollett:2013pwa,Kawasaki:2017bqm}.  Light vector bosons having feeble interactions with the SM sector that we consider in this work are one such example. When these vector bosons decay, energy injected into the cosmological plasma can significantly disrupt the predictions of the standard BBN and such effects can be important for a vector boson with mass above several MeV and having a lifetime $\tau \gtrsim 10^4$ s.

As mentioned before, we consider a scenario where the long-lived massive vector boson has a nonvanishing number density in the early Universe. Electromagnetic energy -- photons, electrons, and positrons are produced directly as well as indirectly (through intermediate muons and pions) from the decays of these vector bosons, which interact
with the plasma before reacting with the light nuclei~\cite{Kawasaki:2004qu,Pospelov:2010cw}.  This late-time high-energetic electromagnetic particle injections into the thermal plasma induce an electromagnetic cascade involving the following processes:
\begin{enumerate}
\item Double photon pair creation: $\gamma \gamma_{BG}\to e^-e^+$, 
\item Photon-photon scattering: $\gamma \gamma_{BG}\to \gamma\gamma$,
\item Bethe-Heitler pair creation: $\gamma N\to e^-e^+ N$, with $N\in (^1\textrm{H},^4\textrm{He})$,
\item Compton scattering: $\gamma e^-_{BG}\to \gamma e^-$, and
\item Inverse Compton scattering: $e^\mp \gamma_{BG}\to e^\mp \gamma$, 
\end{enumerate}
here ``$BG$'' refers to the particle  in the thermal bath. Other processes can be safely neglected due to their small number densities. 

For photons with energies above $E^{e^-e^+}_\textrm{th}\approx \frac{m^2_e}{22\;T}\approx 2 \textrm{MeV}\;\frac{6 \textrm{keV}}{T}$ ($T$ is the temperature of the photon)~\cite{Protheroe:1994dt,Kawasaki:1994sc}, which is the  threshold for double photon pair creation, this process is the most efficient compared to the rest, and therefore, rapidly depletes the high-energy photons. Photodisintegration then can only occur if $E^{e^-e^+}_\textrm{th}$ is above the threshold for the various disintegration reactions.

Even for initial energies orders of magnitude above the MeV scale thresholds for photodisintegration, the fraction of
energy available for photodisintegration is tiny until the background temperature falls below $T\lesssim 10$ keV, which corresponds to $t\sim 10^4$ sec. Since $E^{e^-e^+}_\textrm{th}\propto T^{-1}$, photodisintegration takes place for small enough temperatures,  in particular, 
\begin{enumerate}
\item $\textrm{D}$-disintegration with $E^\textrm{D}_\textrm{th} \approx 2.22 \, \textrm{MeV}$: $T \lesssim 5.34 \, \textrm{keV}$\;, 
\item  ${}^3\textrm{H}$-disintegration with $E^{{}^3\textrm{H}}_\textrm{th} \approx 6.26 \, \textrm{MeV}$: $T \lesssim 1.90 \, \textrm{keV}$\;,
\item  ${}^3\textrm{He}$-disintegration with $E^{{}^3\textrm{He}}_\textrm{th} \approx 5.49 \, \textrm{MeV}$: $T \lesssim 2.16 \, \textrm{keV}$\;,
\item  ${}^4\textrm{He}$-disintegration with $E^{{}^4\textrm{He}}_\textrm{th} \approx 19.81 \, \textrm{MeV}$: $T \lesssim 0.60 \, \textrm{keV}$\;,
\item  ${}^6\textrm{Li}$-disintegration with $E^{{}^6\textrm{Li}}_\textrm{th} \approx 3.70 \, \textrm{MeV}$: $T \lesssim 3.21 \, \textrm{keV}$\;,
\item  ${}^7\textrm{Li}$-disintegration with $E^{{}^7\textrm{Li}}_\textrm{th} \approx 2.47 \, \textrm{MeV}$: $T \lesssim 4.81 \, \textrm{keV}$\;, and
\item  ${}^7\textrm{Be}$-disintegration with $E^{{}^7\textrm{Be}}_\textrm{th} \approx 1.59 \, \textrm{MeV}$: $T \lesssim 7.48 \, \textrm{keV}$\;.
\end{enumerate}
This shows why photodisintegration is only possible
at a late-time and is adequate for relatively low temperatures where BBN has already finished~\cite{Kawasaki:1994sc,Kawasaki:2004qu,Pospelov:2010cw,Poulin:2015woa,Poulin:2015opa,Hufnagel:2018bjp,Forestell:2018txr,Coffey:2020oir}. Owing to the different times involved, the usual BBN and the subsequent photodisintegration reactions factorize. Hence, it is possible to calculate the
nuclear abundances due to nucleosynthesis first and then consider the changes in the abundance as a consequence of photodisintegration.

The electromagnetic cascade spectra of photons and electrons (positron will be the same) evolve according to the following Boltzmann equations: 
 \begin{align}
&\frac{d\mathcal{N}_a}{dt}(E)=- \Gamma_a(E)\mathcal{N}_a(E)+\mathcal{S}_a(E);
\;\;\; \mathcal{N}_a\equiv \frac{dn_a}{dE}, \;\; a= \gamma, e, \label{numberdensity}
 \end{align}
where $\mathcal{N}_a$ the differential number density per unit energy of photons or electrons, $\Gamma_a$ is the net damping rate for species $a$ at energy $E$ to lower energies, and $\mathcal{S}_a$ is the injection rate from all sources at energy $E$. Compared to the Hubble rate and the effective photodisintegration rates with light nuclei, the damping and transfer reactions are generally fast; hence,  the quasistatic limit, $\frac{d\mathcal{N}_a}{dt}\to 0$ is a good approximation~\cite{Cyburt:2002uv}, which leads to 
\begin{align}
\mathcal{N}_a=\frac{\mathcal{S}_a(E)}{\Gamma_a(E)}.    
\end{align}
Here, both $\mathcal{S}_a(E)$ and $\Gamma_a(E)$ vary adiabatically as functions of time.

The source terms in Eq.~\eqref{numberdensity} receive
contributions from direct injection and also  from transfer reactions moving energy from higher up in the cascade down to $E$. Therefore, it takes the following form, 
\begin{align}
\mathcal{S}_a=R \frac{dN_a}{dE}+ \sum_b \int_E^{E_X} dE^\prime K_{ab}(E,E^\prime) \mathcal{N}_b(E^\prime),     
\end{align}
where $\frac{dN_a}{dE}$ is the primary energy spectrum per injection of  photon or electron/positron collected in Appendix \ref{appendix-B}, and $R$ is the injection rate from the decay species $V$: $R=e^{-t/\tau} n^0_V/\tau$, where $\tau$ is the lifetime, and $n^0_V$ is the number density of $V$ in the absence of its decay. Moreover, in the second term, $E_X$ is the maximum energy in the cascade and  $K_{ab}(E,E^\prime)$ is the transfer kernel, which describes reactions $b(E^\prime)+X_{BG}\to a(E)+X^\prime_{BG}$ within the cascade.

Finally, the effect of photodisintegration on the abundances of light elements is dictated by the following Boltzmann equations:   
\begin{align}
\frac{dY_A}{dt}=\sum_i Y_i \int^\infty_0 dE_\gamma \mathcal{N}_\gamma(E_\gamma) \sigma_{\gamma +i\to A}(E_\gamma)  -
 Y_A
\sum_f \int^\infty_0 dE_\gamma \mathcal{N}_\gamma(E_\gamma) \sigma_{\gamma +A\to f}(E_\gamma), \label{eq:Boltzmann}
\end{align}
where $\mathcal{N}_\gamma(E_\gamma)$ are the photon spectra calculated above; $A$ and the sums run over the relevant
isotopes. The isotope number density normalized to entropy density is defined as before, $Y_A= n_A/s$. 
To solve the equations above, we utilize the public code \texttt{ACROPOLIS}~\cite{Depta:2020mhj,Depta:2020zbh,Hufnagel:2018bjp}. We have created a general vector boson model file with the resulting primary photon and electron/positron spectra from vector boson decay for \texttt{ACROPOLIS} as described in the Appendix \ref{appendix-A}.
The initial conditions after the BBN are set to the outputs of \texttt{AlterBBN} utilizing the mean, high, and low values of the nuclear reaction rates~\cite{Arbey:2011nf,Arbey:2018zfh}. We will \emph{conservatively} solve Eq.~\eqref{eq:Boltzmann} down till the epoch of matter-radiation equality at the dawn of structure formation at $t \sim 2\times 10^{12}$\,s. After solving Eq.~\eqref{eq:Boltzmann} with \texttt{ACROPOLIS}, we estimate the ``theoretical'' error in the light element abundances by conservatively taking
%%%
\begin{equation}
    \sigma_{Y_A} = \max\left[|Y_A(\textrm{high}) - Y_A(\textrm{mean})|,|Y_A(\textrm{low}) - Y_A(\textrm{mean})|\right],
    \label{eq:th_error}
\end{equation}
%%%
where $Y_A(\textrm{mean})$, $Y_A(\textrm{high})$, and $Y_A(\textrm{low})$ are respectively the outputs of \texttt{ACROPOLIS} using the initial conditions produced by \texttt{AlterBBN} with mean, high, and low values of the nuclear reaction rates.

To compare with the observed light element abundances, we consider the Helium-4 mass fraction, deuterium abundances from the Particle Data Group~\cite{ParticleDataGroup:2020ssz}, and Helium-3 abundance from Ref.~\cite{Bania:2002yj}
\begin{align}
&Y_p=0.245\pm 0.003,\;\;\;
&\frac{n_{\rm D}}{n_{\rm H}}=(2.547\pm 0.025)\times 10^{-5},\;\;\;
&\frac{n_{^3{\rm He}}}{n_{\rm H}}=(1.1\pm 0.2)\times 10^{-5}. 
\label{eq:exp_error}
\end{align}
For the exclusion at 95\% CL, we consider each of the elements separately, summing the errors from Eqs.~\eqref{eq:th_error} and \eqref{eq:exp_error} in quadrature.

%%%%%%%%%%%%%%%%%%%%%%%%%%%%%%%%%%%%%%%%%%%%%%%
%%%%%%%%%%%%%%%%%%%%%%%%%%%%%%%%%%%%%%%%%%%%%%%
\section{Application to the Dark Photon Model}
\label{sec:dark_photon}

As an application, we will consider light vector boson $V$ as a dark photon that couples to the SM only through kinetic mixing Eq.~\eqref{eq:kinetic_mixing}. We focus on the scenario with $\epsilon \lesssim 10^{-9}$ in which the dark photons are never in thermal equilibrium with the SM sector.\footnote{For the constraints on dark photon with $\epsilon \gtrsim 10^{-9}$, see, e.g., the review article~\cite{Fabbrichesi:2020wbt}.} We will first determine the production-independent BBN photodisintegration constraints on the dark photon energy density in Section \ref{subsec:BBN_dark_photon}. Then, we will review the dark photon production from the freeze-in mechanism in Section \ref{subsec:freeze-in}. In Section \ref{sec:constraints_dark_photon}, we will include both gravitational and freeze-in contributions and show that the former is dominant when $H_I \gtrsim 10^8$ GeV leading to stringent constraints on the parameter space of the dark photon.

%%%%%%%%%%%%%%%%%%%%%%%%%%%%%%%%%%%%%%%%%%%%%%%%%%%%%%
\subsection{BBN photodisintegration constraints on the dark photon}
\label{subsec:BBN_dark_photon}

\begin{figure}[ht!] 
    \centering
    \includegraphics[width=0.75\textwidth]{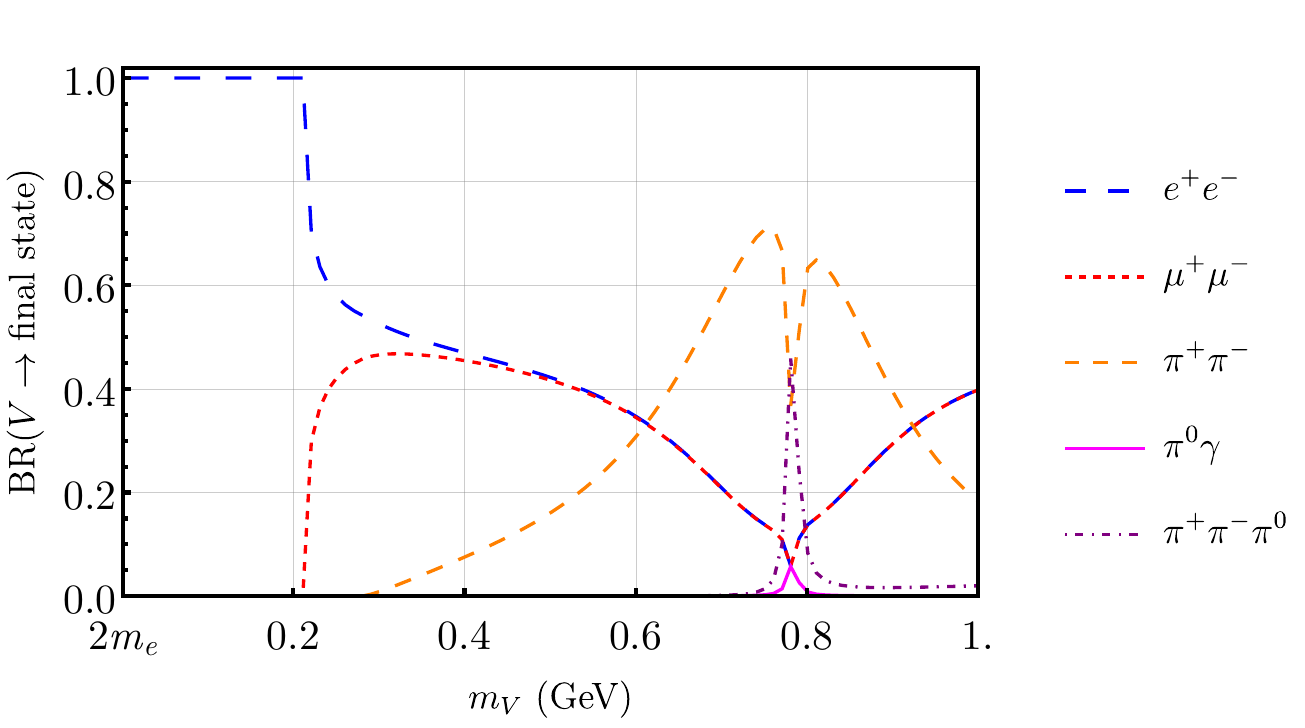}
    \caption{Branching ratios for the dark photon decay to different final states.
    %\textcolor{red}{[The branching ratio should start from $2m_e$ but I guess hard to see on the plot. We can also make a log scale.]} \textcolor{black}{yes, it starts at 2$m_e$, we can by hand change the tick from 0 to 2$m_e$?}
    } \label{fig:BR}
\end{figure}

We will first derive the production-independent BBN photodisintegration constraints on the dark photon energy density $m_V Y_V$ as function of mass $m_V$ and lifetime $\tau$ of the dark photon. We obtain the constraints by inputting the decay branching ratios of dark photon as shown in Figure \ref{fig:BR} (which is calculated using the formulas in Section \ref{decaywidths}) to the general light vector model for \texttt{ACROPOLIS} as described in Appendix \ref{appendix-A} and then we carry out the $\chi^2$ analysis taking into account both the theoretical and experimental errors given in Eqs.~\eqref{eq:th_error} and \eqref{eq:exp_error}.

\begin{figure}[!ht] 
    \centering
    \includegraphics[width=\textwidth]{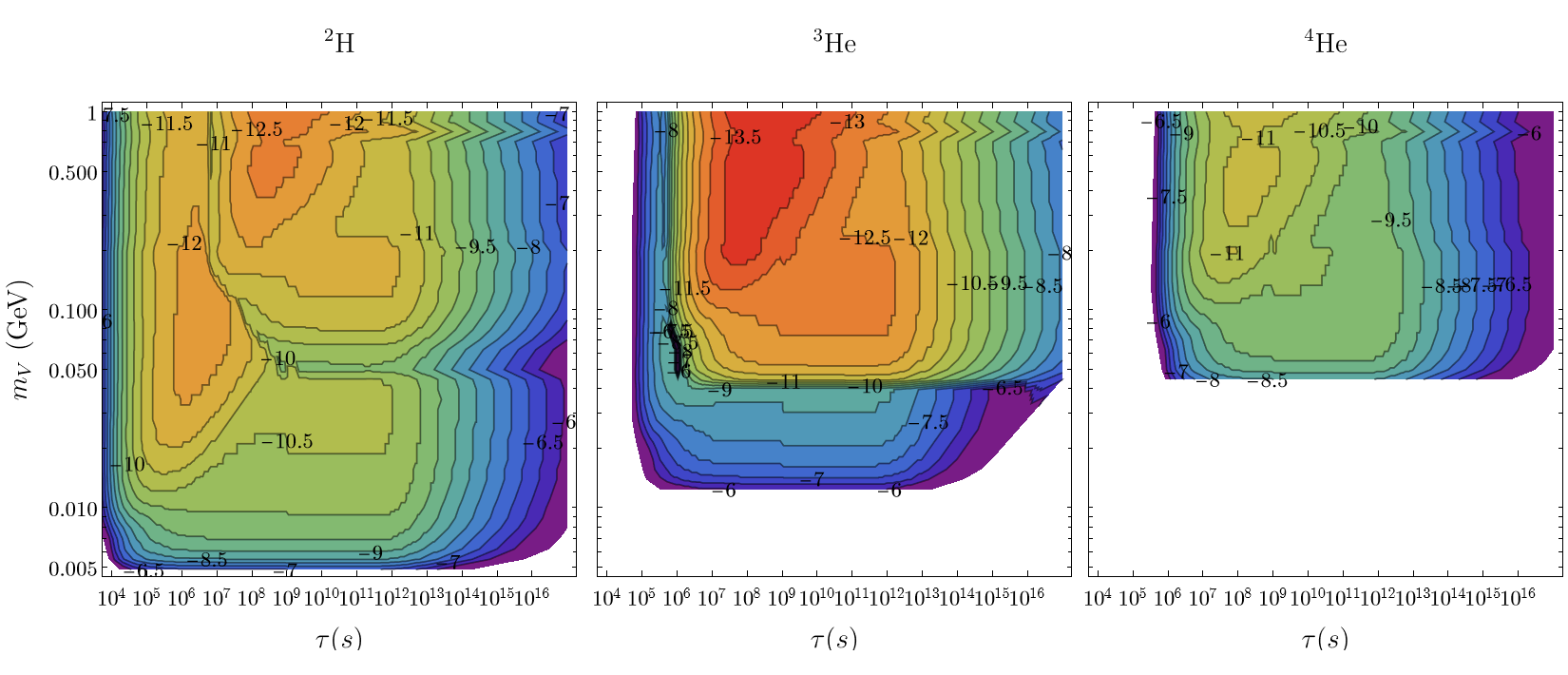}
    \caption{BBN photodisintegration constraints on mass times abundance of dark photons from measurements of primordial abundances of ${}^2 \text{H}, {}^3\text{He}$, and ${}^4\text{He}$ quoted in Eq.~\eqref{eq:exp_error}. The contours represent the upper bounds in $\log_{10}({m_V Y_V / \textrm{GeV}})$. \label{fig:BBNconstraints}
    %where $m_V$ is in GeV. 
    }
\end{figure}

In Figure \ref{fig:BBNconstraints}, we show the upper bound on $m_V Y_V$ at the 95\% CL on the plane of $m_V$ and $\tau$. 
\textcolor{black}{For very short lifetime $\tau$ of $V$, as we discussed in the previous section, the thermal bath photons are still energetic to interact with high energy photons from decays of $V$ to create $e^- e^+$, quickly depleting the energy of these photons, leaving very little energetic photons to photodisintegrate the light elements.}
We see that the constraints start to be relevant first (from the smallest $\tau$ and $m_V$) for $^2$H and then followed by $^3$He and $^4$He due to the increasing threshold energies for photodisintegration for these elements. 
\textcolor{black}{As $\tau$ increases further, the constraints again become weaker beyond $\tau \sim 10^{12}$ s since most of the electromagnetic energy is injected into the cosmic plasma after matter-radiation equality where we stop the evaluation since the formalism used here is not adequate to describe the photodisintegration after this epoch. Hence the limits we have obtained in this regime should be considered the most conservative ones. The detail features of the constraints can be traced to the available decay channels and branching ratios which are model-dependent.} 
We do not show the constraint for $m _V Y_V > 0.9\times 10^{-6}$ GeV since this is not consistent with the BBN as discussed in the previous section, cf. Eq. \eqref{eq:BBN_constraint}. Although the constraint from $^3$He is the strongest in part of the parameter space, we caution that inferring the primordial $^3$He abundance~\cite{Bania:2002yj} is not a settled issue since stellar nucleosynthesis models for $^3$He are in conflict with observations~\cite{Olive:1996tt}. 

%%%%%%%%%%%%%%%%%%%%%%%%%%%%%%%%%%%%%%%%%%%%%%%
%%%%%%%%%%%%%%%%%%%%%%%%%%%%%%%%%%%%%%%%%%%%%%%
\subsection{Freeze-in production}
\label{subsec:freeze-in}

\begin{figure}[!ht] 
    \centering
    \subfloat[Total decay widths \label{dwvsmass}]{
        \includegraphics[width=0.48\textwidth]{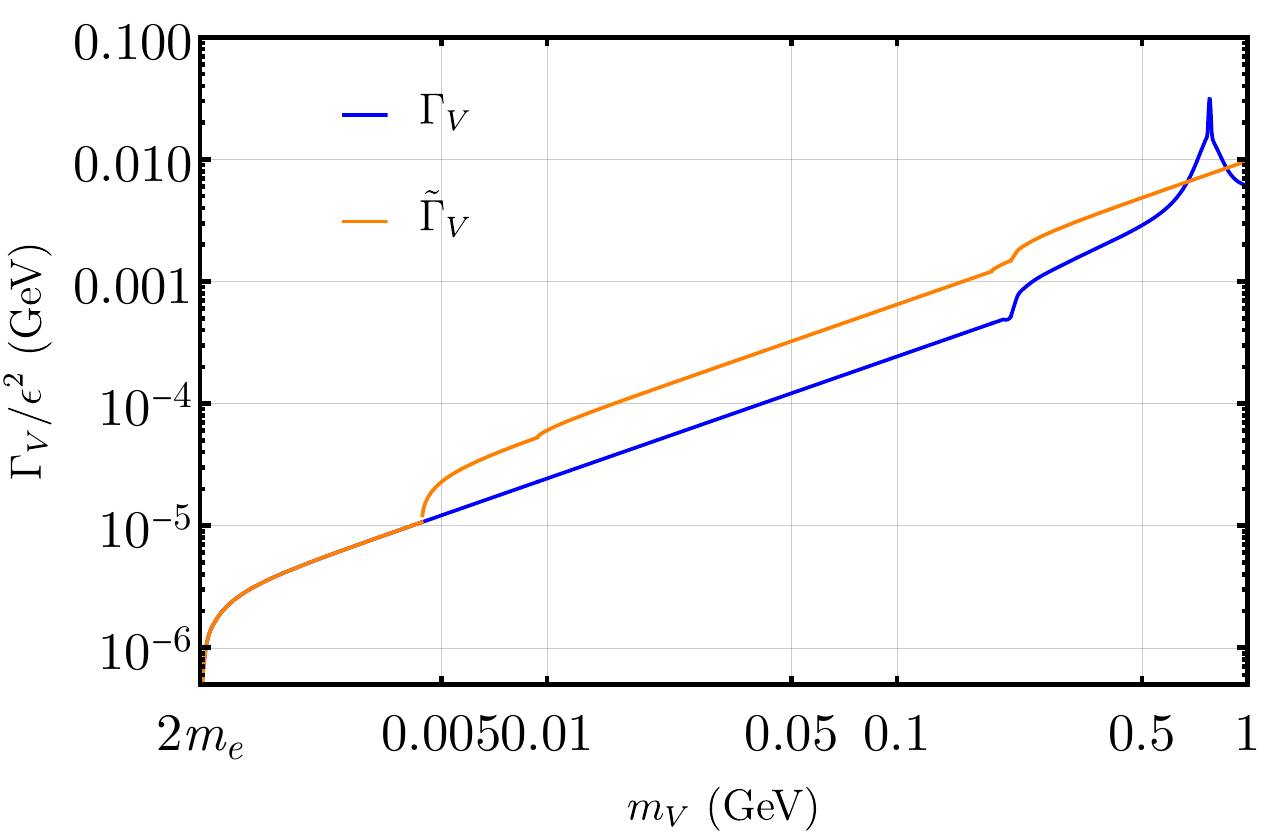}
    }
    \subfloat[Freeze-in yield \label{Yfvsmass}]{
        \includegraphics[width=0.48\textwidth]{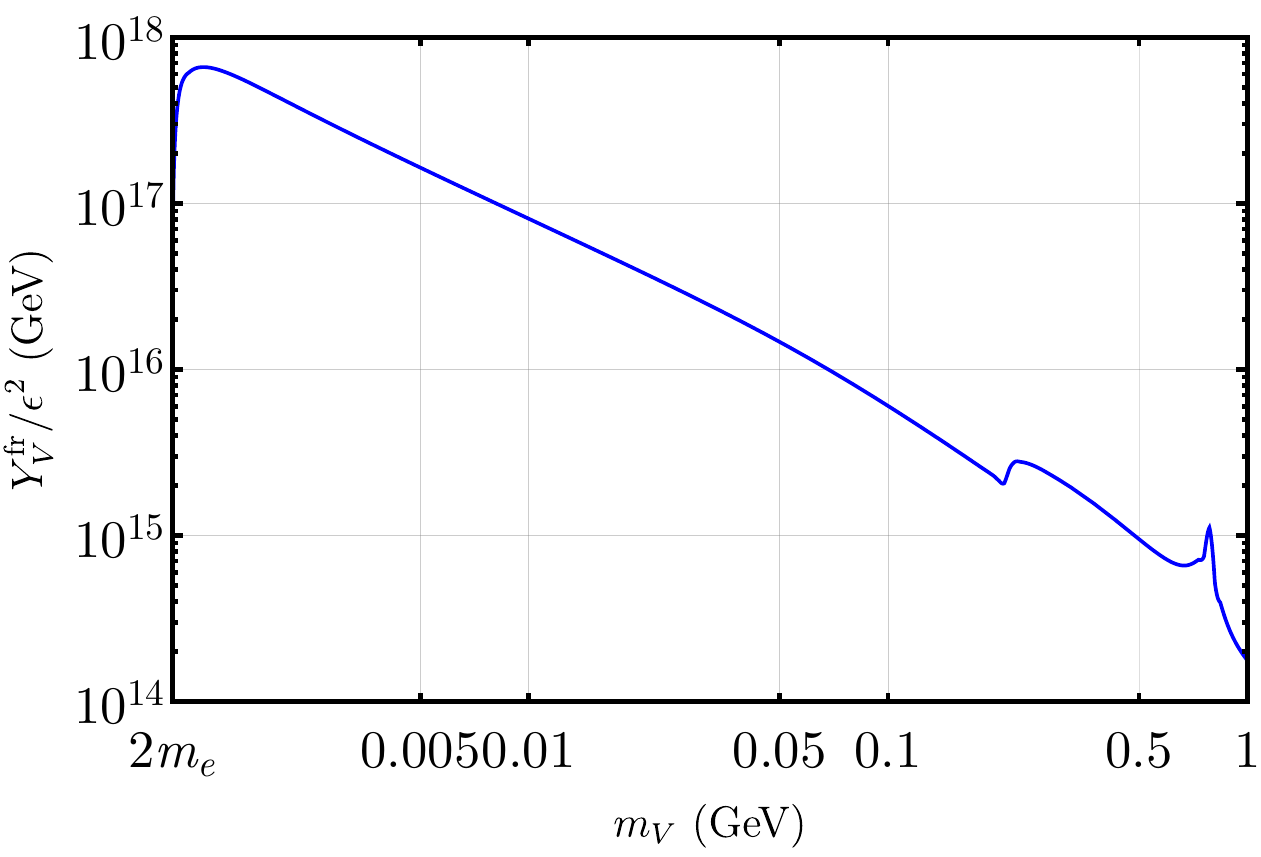}
    }
    \caption{\emph{Left panel:} Total decay width appearing in Eq.~\eqref{YfQCD}. \emph{Right panel:} Freeze-in yield using Eq.~\eqref{YfQCD}. 
    %\textcolor{red}{[For me $\Gamma_V$ seems to start from $2m_e$. The question is why $\tilde \Gamma_V$ does not start at $2m_e$?]}
    }
    \label{fig:freeze-in}
\end{figure}

The main freeze-in production of  vector bosons occurs via inverse decay of SM
charged fermions (both quarks and leptons) as discussed in section \ref{sec:vector_model}. Assuming the latest period of radiation domination once had a temperature $T \gg m_V$ with no significant entropy injection for $T \lesssim m_V$, the evolution of the freeze-in yield with respect to the temperature can be expressed as~\cite{Kolb:1990vq}
\begin{align}
    s H T \frac{dY^{\textrm{fr}}_V}{dT} = -C_f, \label{Yfreezein}
\end{align}
where $C_f$ is the standard collision term and is a function of the phase space distribution of particles involved in the interaction. Assuming the SM particles are in thermal equilibrium following Maxwell-Boltzmann statistics and neglecting thermal corrections and the distribution of vector bosons (it is much smaller than the equilibrium one),\footnote{Using thermal corrections and Fermi-Dirac statistics does not change the result significantly~\cite{Fradette:2014sza}.} the collision term can be expressed as~\cite{Gondolo:1990dk, Edsjo:1997bg}
\begin{align}
    C_f = \frac{3}{2\pi^2} \Gamma_V m_V^2 T K_1\left(\frac{m_V}{T}\right),
\end{align}
where $K_1$ is the $1$st order modified Bessel function of the second kind, and the decay width $\Gamma_V$ is the sum of available decay channels discussed in section \ref{decaywidths}. Dividing the production into temperatures below and above the QCD phase transition temperature $T_{\textrm{QCD}} \simeq 157$ MeV, we can write the solution of Eq.~\eqref{Yfreezein} as
\begin{align}
    Y^{\textrm{fr}}_V = \frac{3}{2\pi^2} m_V^2 \left[ \Gamma_V \int_{T_f}^{T_{\textrm{QCD}}} dT \frac{K_1(m_V/T)}{s H} +  \tilde{\Gamma}_V \int_{T_{\textrm{QCD}}}^{T_i} dT \frac{K_1(m_V/T)}{s H}\right], \label{YfQCD}
\end{align}
where $\Gamma_V$ is the total decay to leptons and hadronic channels discussed in Section \ref{decaywidths}, and $\tilde{\Gamma}_V$ is the vector decay width into perturbative quark and lepton final states, where one can use Eq.~\eqref{decayw} for leptons, and use the same equation but include an additional factor of $3$ for quarks. The total decay widths $\Gamma_V$ and $\tilde\Gamma_V$ as a function of $m_V$ is shown on the left plot of Figure \ref{fig:freeze-in}. Since the light vector bosons do not dominate the cosmic energy density, it is a good assumption to use $H(T)=\sqrt{\frac{\pi^2}{90}g_\star(T)} \frac{T^2}{M_P}$ assuming the SM relativistic degrees of freedom $g_\star(T)$ that we have obtained from Ref.~\cite{Drees:2015exa}. The result of the numerical integration of Eq.~\eqref{YfQCD} is shown in the right plot of Figure \ref{fig:freeze-in}. Note that the final result does not depend on the precise value of $T_f$ in Eq.~\eqref{YfQCD}, since the production is very suppressed for $T_f \ll m_V$, nor does it depend sensitively on the initial temperature as long as $T_i \gg m_V$.

%%%%%%%%%%%%%%%%%%

%%%%%%%%%%%%%%%%%%%%%%%%%%%%%%%%%%%%%%%%%%%%%%%
%%%%%%%%%%%%%%%%%%%%%%%%%%%%%%%%%%%%%%%%%%%%%%%
\subsection{Constraints on gravitationally produced dark photons}
\label{sec:constraints_dark_photon}

\begin{figure}[!ht] 
    \centering
    \subfloat[ \label{YV1}]{
        \includegraphics[width=1.03\textwidth]{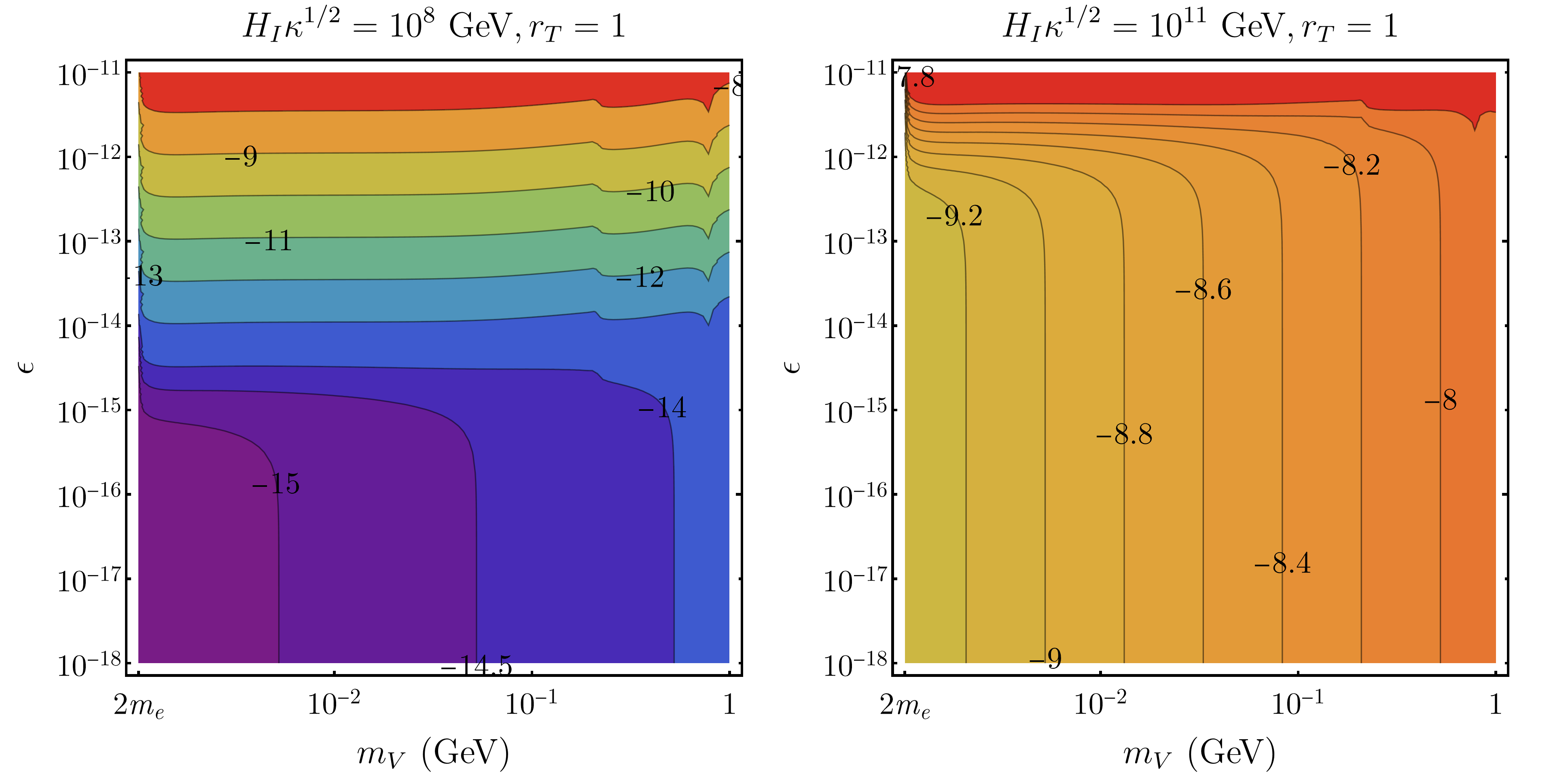}
    }\\
    \subfloat[\label{YV2}]{
        \includegraphics[width=1.03\textwidth]{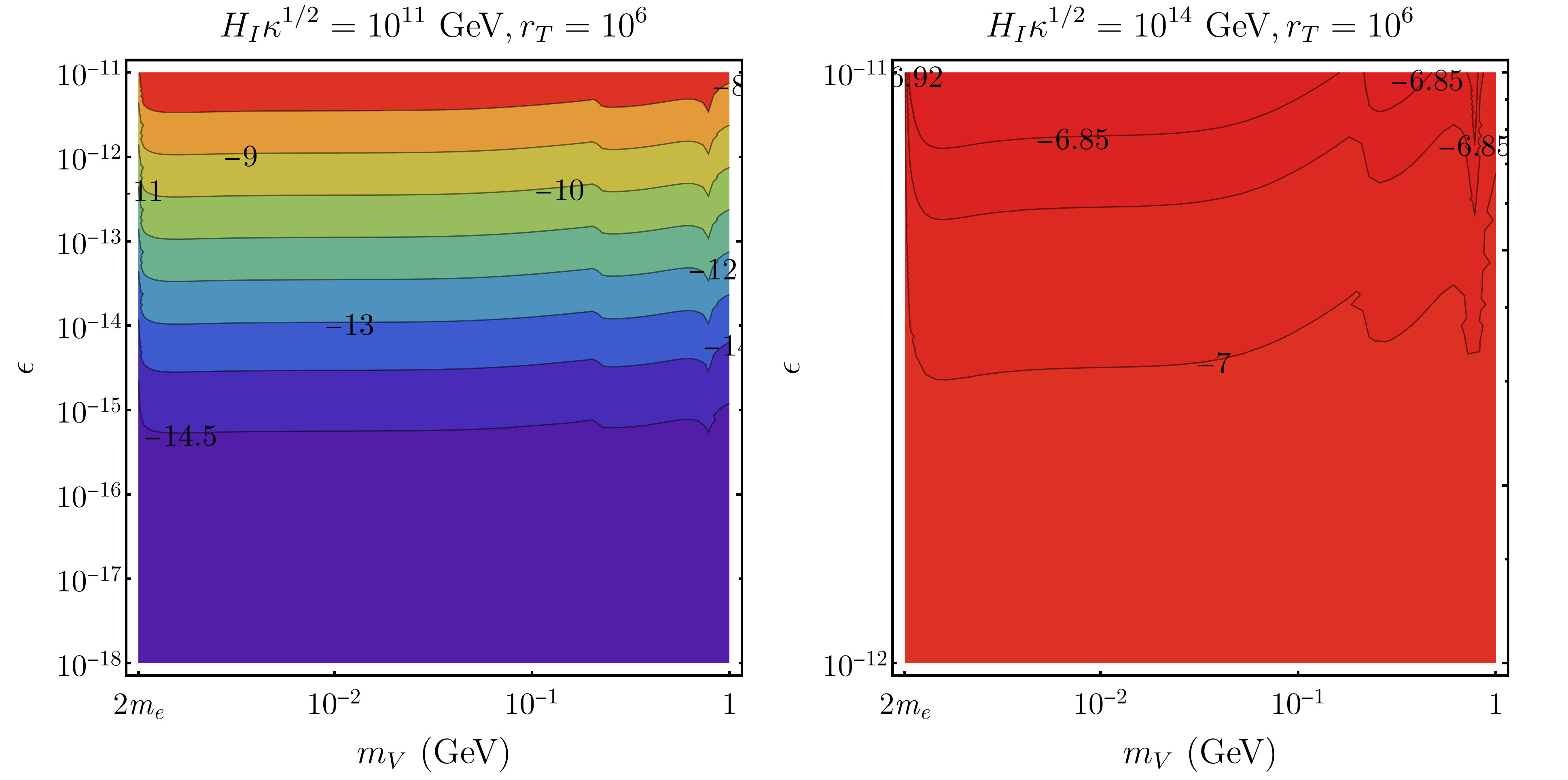}
    }
    \caption{$\log_{10}({m_V Y_V / \textrm{GeV})}$ contours of the dark photons produced from both gravitational and freeze-in mechanisms in (a) high reheating case $r_T = 1$, and (b) low reheating case $r_T = 10^6$.    
    %\textcolor{red}{[Can we remove 0.005, 0.05, 0.5 labels in the $x$-axis? It will make them cleaner and easier to read.]}
    }
    \label{fig:total_YV}
\end{figure}

As shown in Figure~\ref{fig:total_YV}, for high reheating scenario $r_T = 1$, the dark photon is completely dominated by gravitational production for $H_I = 10^{11}$ GeV (i.e. roughly independent of $\epsilon$) while for $H_I = 10^{8}$ GeV, freeze-in production starts to be comparable to gravitational production starting from $\epsilon \gtrsim 10^{-15}$ where the dependence on $\epsilon$ becomes apparent. 
The situation for low reheating scenario $r_T = 10^6$ is similar but the freeze-in production becomes relevant already at $H_I \lesssim 10^{11}$ GeV due to additional suppression in gravitational production as we can see in Eq.~\eqref{eq:gravyield} or \eqref{eq:Ygrav_2cases}. \textcolor{black}{In the gravitational production dominated regime, from Eq.~\eqref{eq:Ygrav_2cases}, $m_V Y_V$ depends on $m_V$ for high reheating scenario, but becomes independent of $m_V$ for low reheating case. In Figure \ref{fig:total_YV}, this is evident by vertical contours in the former case but horizontal contours in the latter. Note that in the bottom right figure, the $\epsilon$ axis has been shown for $10^{-12} \leq \epsilon \leq 10^{-11}$ to make the contours visible (variations due to freeze-in contributions), and for $\epsilon < 10^{-12}$, $m_V Y_V \simeq 10^{-7}$, as can be verified from Eq.~\eqref{eq:Ygrav_2cases}.}

\begin{figure}[!ht] 
    \centering
    \includegraphics[width=1.035\textwidth]{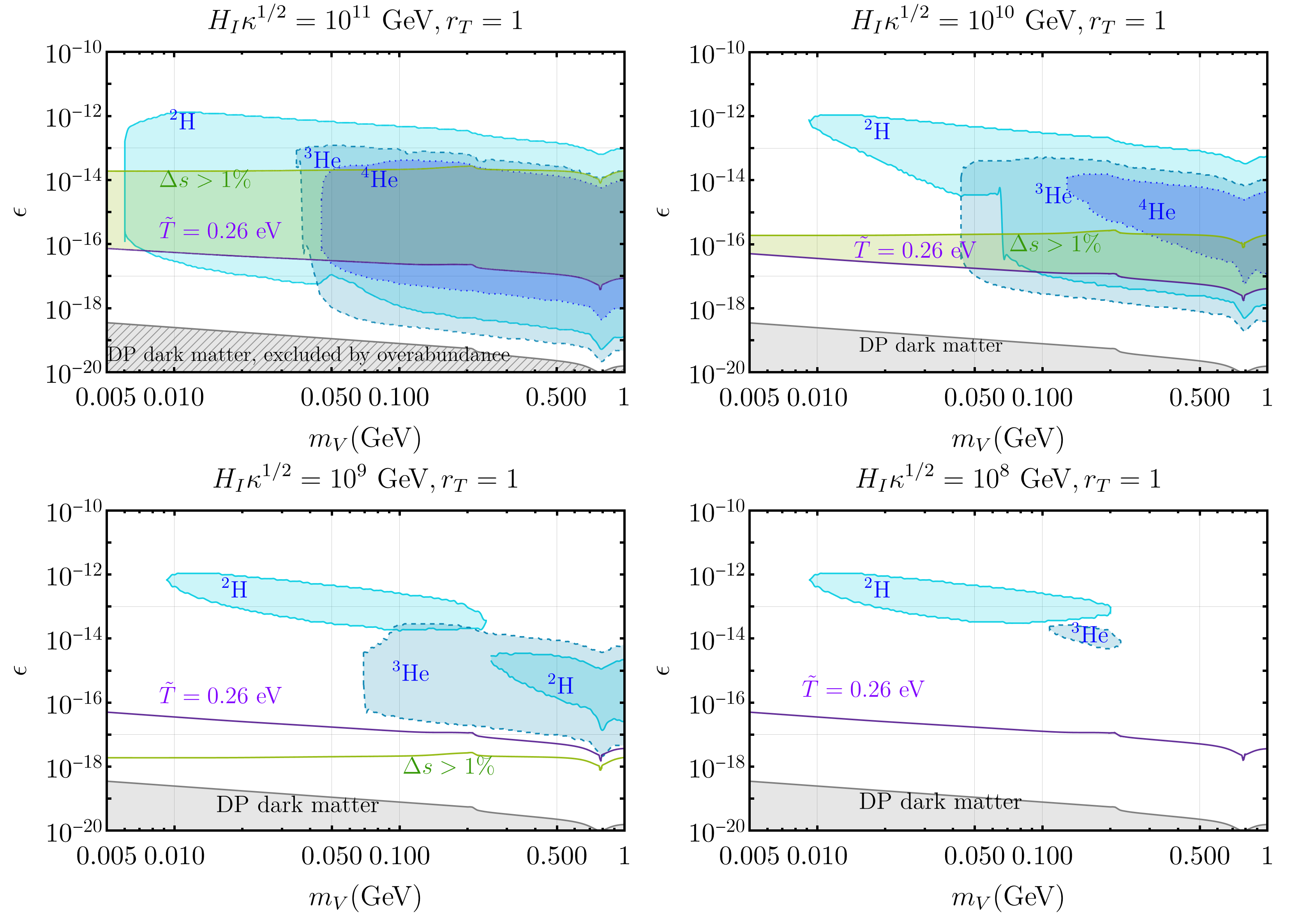}
    \caption{
    The shaded regions delimited by solid, dashed, and dotted lines are ruled out at 95\% CL due to the BBN photodisintegration effects from the measurement of primordial abundance of $^2$H, $^3$He, and $^4$He, respectively. The green shaded area ($\Delta s > 1$\%) denote the region where the entropy injection from the dark photon decay after the BBN but before the CMB (after the purple line $\tilde T = 0.26$ eV) is larger than 1\%. In the shaded area below the gray line (DP dark matter), the dark photon has a lifetime longer than the age of the Universe and is a dark matter.
    }\label{fig:high_rT}
\end{figure}

\begin{figure}[!ht] 
    \centering
    \includegraphics[width=1.035\textwidth]{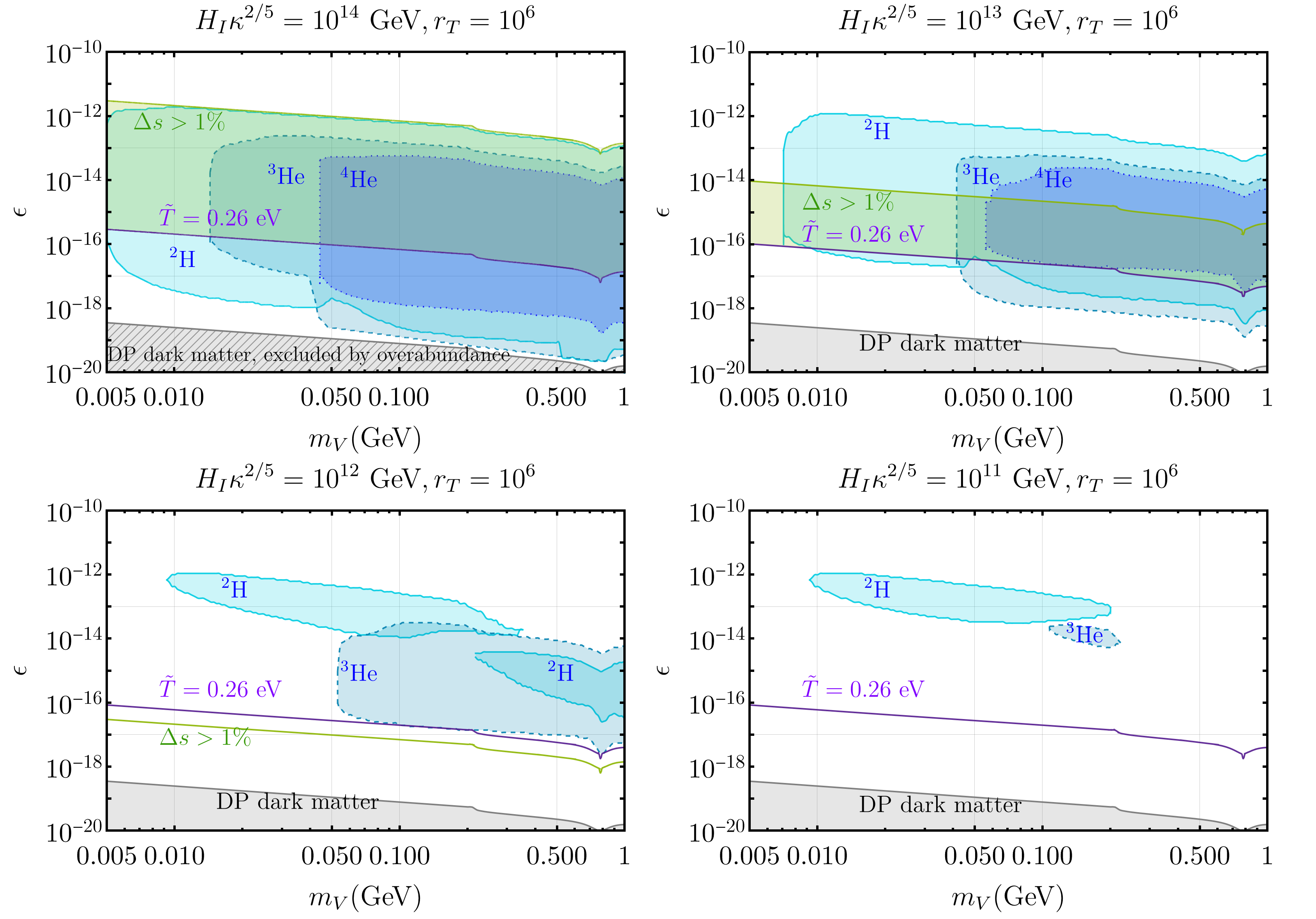}
    \caption{The same as Figure \ref{fig:high_rT} but for low reheating scenario $r_T = 10^6$.   
    }\label{fig:low_rT}
\end{figure}

In Figure \ref{fig:high_rT}, we show the BBN photodisintegration constraints on the parameter space (kinetic mixing $\epsilon$ and mass $m_V$) of the dark photon for high reheating scenario $r_T = 1$.
%---------
The shaded regions delimited by solid, dashed, and dotted lines are ruled out 95\% CL by the measurement of the primordial abundance of $^2$H, $^3$He, and $^4$He, respectively. The shaded green area ($\Delta s > 1$ \%) is the area where the entropy injection after the BBN but before the CMB (above the purple line with $\tilde T = 0.26$ eV) is greater than 1\% from imposing Eqs.~\eqref{eq:decay_bound} and \eqref{eq:decay_condition}. In the shaded area below the gray line (DP dark matter), the dark photon has a lifetime longer than the age of the Universe and is a dark matter. For $H_I \kappa^{1/2} = 10^{11}$ GeV, the dark matter parameter space $m_V > 2m_e$ is excluded since it exceeds the observed dark matter energy density [see Eq.~\eqref{eq:viable_DM}]. For lower $H_I$, the dark photon can be a metastable dark matter subject to astrophysical constraints. For the dark photon we consider in this work to be a dark matter candidate, its lifetime needs to be quite large. Since it eventually decays to electron-positron and/or photons, its lifetime is constrained by observations of the galactic and extra-galactic diffuse X-ray or gamma-ray background. Consequently, data from satellites such as HEAO-1, INTEGRAL, COMPTEL, EGRET, and Fermi-LAT telescopes can be used to put a constraint on the lifetime as a function of the dark photon mass. One important constraint is the dark photon decay to electron-positron with final-state radiation. Bound on the dark matter decay lifetime for this process in the mass range 1 MeV - 1 GeV is typically of order $\tau \gtrsim \mathcal{O}(10^{25})$ s~\cite{Essig:2013goa}.\footnote{\textcolor{black}{For purely gravitationally produced dark matter, the constraint from the Lyman-$\alpha$ forrest is very weak~\cite{Ballesteros:2020adh}, for example, the lower bound on the mass for the gravitationally produced scalar dark matter corresponds to $m> 0.34$ meV~\cite{Garcia:2022vwm}. For a vector dark matter, we expect a bound on its mass from Lyman-$\alpha$ constraint of similar order.}}.
One notices that when $H_I$ is smaller, the constraints get weaker since the gravitational production is less efficient. For $H_I \kappa^{1/2} = 10^8$ GeV, the freeze-in production of the dark photon becomes comparable to the gravitational production and starts to be dominant for $\epsilon \gtrsim 10^{-15}$ and the BBN constraints from $^2$H and $^3$He in the lower right plot of Figure \ref{fig:high_rT} is essentially due to the freeze-in production of the dark photon. Furthermore, in Figure \ref{fig:low_rT}, we show the constraints for low reheating scenario $r_T = 10^6$. The main difference is that the gravitational production is more suppressed and similar constraints with Figure \ref{fig:high_rT} are obtained but with larger $H_I$.
%---------
%--------------

\textcolor{black}{ Here we briefly summarize the constraints on the dark photon in the $(m_V,\epsilon)$ plane arising from accelerators and other experiments (for a recent review, see, for example, Ref.~\cite{Fabbrichesi:2020wbt}).  We restrict our discussion to the sub-GeV mass range, which is of our interest.   LHCb~\cite{LHCb:2019vmc} searches for dark photons produced in $pp$ collisions  at a center-of-mass energy of 13 TeV and looking for decays of the type $V\to \mu^+\mu^-$. In the mass range $m_V=[214-740]$ MeV, it constraints the kinetic mixing in the range $\epsilon<[10^{-4}, 10^{-3}]$.  Exclusion limits of A1 Collaboration~\cite{Merkel:2014avp} (searching for electron-positron pair-production from $V$)  in the mass range $m_V=[40-300]$ MeV is  down to $\epsilon\sim 8\times 10^{-4}$. Dark photon production in the $\pi^0\to \gamma V\to \gamma e^+e^-$ decay is searched for at the NA48/2~\cite{NA482:2015wmo} experiment, which restricts the mixing parameter as low as $\epsilon\sim 4\times 10^{-4}$ for $m_V=[9-70]$ MeV.  Apart from these  collider/fixed target experiments, dark photon mass in the window $m_V=[1-30]$ MeV is constrained to be $\epsilon \lesssim 10^{-5}$ from beam dump experiment E141~\cite{Riordan:1987aw}.  Other beam dump experiments such as $\nu$-cal~\cite{Blumlein:2011mv,Blumlein:2013cua}, CHARM~\cite{Gninenko:2012eq}, and E137~\cite{Batell:2014mga,Marsicano:2018krp} rule out a large portion of the parameter space in the entire 1 MeV to 1 GeV mass range for couplings in between $\epsilon\sim [10^{-7}-10^{-4}]$. Finally, bounds from supernovae ~\cite{Chang:2016ntp} rules out $\epsilon\sim [10^{-9}-10^{-7}]$ in the $m_V=[1-200]$ MeV  mass regime. All these experiments together put very tight constraints on the kinetic mixing parameter with $\epsilon\gtrsim 10^{-9}$ in the range 1 MeV to 1 GeV. However, all these constraints disappear for $\epsilon\lesssim 10^{-9}$. Our work shows a new parameter space in the same mass range with much lower kinetic mixing  $\epsilon\sim 10^{-19}-10^{-12}$ that is tightly constrained from BBN bounds, which depends on the Hubble scale at the end of inflation $H_I$ as discussed above. }

%%%%%%%%%%%%%%%%%%%%%%%%%%%%%%%%%%%%%%%%
%%%%%%%%%%%%%%%%%%%%%%%%%%%%%%%%%%%%%%%%
\section{Conclusions and Future Directions}
\label{sec:conclusion_future}

In this work, we have considered a MeV to GeV scale new vector boson $V$, which couples feebly to the SM, and hence the particles never reach thermal equilibrium. Gravitational production during the epoch of inflation and reheating can be significant if the Hubble scale at the end of the inflation $H_I$ is large. For $H_I \gtrsim 10^8\,(10^{11})$ GeV, gravitational production starts to dominate over the freeze-in production for high (low) reheating scenario. For high (low) reheating, if $V$ has a lifetime longer than the age of the Universe $\tau \gtrsim 4\times 10^{17}$\,s, with mass in the range MeV to GeV, it can constitute all the metastable dark matter if $H_I \sim 10^{10}-10^{11}$\,GeV ($H_I \sim 10^{13}$\,GeV). 
Nevertheless, its decays to the SM particles are subject to strong astrophysical constraints, which impose its lifetime to be $\tau \gtrsim \mathcal{O}(10^{25})$ s~\cite{Essig:2013goa}. If $V$ has a lifetime $\tau \gtrsim 10^4$\,s, it decays after the completion of the BBN, and very stringent constraints on the mass and lifetime of $V$ can be derived from the photodisintegration effects on the light element abundances compared to the scenario where only freeze-in production is considered. 

We have created a model file for $V$ including all the primary photon and electron/positron spectra that can be input to \texttt{ACROPOLIS} to determine the photodisintegration constraints on $m_V$ and $\tau$.
As an example, we have investigated a dark photon model that couples to the SM only through kinetic mixing. We are able to rule out a large portion of parameter space of kinetic mixing parameter $\epsilon$ and $m_V$ for $H_I \gtrsim 10^8 (10^{11})$ GeV for high (low) reheating scenario. In summary, if $H_I$ is determined to be $10^{14}$ GeV, the existence of MeV to GeV mass scale dark photon with kinetic mixing $\epsilon \lesssim 10^{-13}$ is essentially completely ruled out.

\vspace{0.3cm} \noindent
Finally, our work can be extended to a few directions:
\begin{itemize}
    \item For the BBN photodisintegration study with $\tau \gtrsim 10^4\,{\rm s}$, our model file can be extended for heavier gauge boson $m_V \gtrsim 1$ GeV.
    
    \item To derive constraints for shorter lifetime $1\,{\rm s}\lesssim \tau \lesssim 10^4\,{\rm s}$ and heavier $m_V$, our work can be extended to study hadrodisintegration effects on the BBN~\cite{Jedamzik:2004er,Kawasaki:2004qu,Pospelov:2010cw}. 
    
    \item For $\tau \gtrsim 10^{13}\,{\rm s}$, there is also constraints from the CMB power and frequency spectra. As shown in Ref.~\cite{Coffey:2020oir}, considering only the freeze-in production, a large parameter space for a light vector boson is already ruled out. Taking into account the gravitational production, stronger constraints are expected even for a relatively small value of $H_I$. 
\end{itemize}

%%%%%%%%%%%%%%%%%%%%%%%%%%%%%%%%%%%%%%%%%%%%%%%
%%%%%%%%%%%%%%%%%%%%%%%%%%%%%%%%%%%%%%%%%%%%%%%
\section*{Acknowledgments}
C.S.F. acknowledges the support by grant 2019/11197-6 and 2022/00404-3 from São Paulo Research Foundation (FAPESP), and grant 301271/2019-4 from National Council for Scientific and Technological Development (CNPq). M.H.R. acknowledges partial support from Maurice C. Holmes and Frances A. Holmes Endowed Fellowship, and U.S. Department of Energy under grant number DE-SC0010296. The authors acknowledge University of Florida Research Computing for providing computational resources, and David Hansen for support. We thank Marco Hufnagel for the correspondence on \texttt{ACROPOLIS}.

\appendix
%%%%%%%%%%%%%%%%%%%%%%%%%%%%%%%%%%%%%%%%
%%%%%%%%%%%%%%%%%%%%%%%%%%%%%%%%%%%%%%%%
\section{Appendix: Decaying Vector Boson Model for \texttt{ACROPOLIS}}\label{appendix-A}

We implement the model of a general light vector boson with mass in the range MeV to GeV for the public code \texttt{ACROPOLIS}~\cite{Depta:2020mhj,Depta:2020zbh,Hufnagel:2018bjp}. Our implementation is available on GitHub at \url{https://github.com/shengfong/lightvectorboson}. 
The primary photon and electron/positron spectra 
from all five possible channels that are relevant for $m_V \leq \textrm{GeV}$ are computed: $V \to e^+ e^-$, $V \to \mu^+ \mu^-$, $V \to \pi^+ \pi^-$, $V \to \pi^0 \gamma$, and $V \to \pi^0 \pi^+ \pi^-$.  We have made appropriate modifications to be able to take into account two monochromatic injection energies: that of electron/positron in $V \to e^+ e^-$ and that of photon in $V \to \pi^0 \gamma$. The model file  \texttt{decay\_vector\_model.py} contains the analytic expressions for the primary photon and electron/positron spectra for $V \to e^+e^-$ and $V \to \pi^0 \gamma$ as well as the final state radiation of photons for $V\to \mu^+\mu^-$ and $V\to \pi^+ \pi^-$. The primary photon and electron/positron spectra for $V\to \mu^+\mu^-$, $V\to \pi^+ \pi^-$, and $V \to \pi^0 \pi^+ \pi^-$ are precalculated numerically (details in Appendix \ref{appendix-B}) and the data is stored in the folder \texttt{spec\_data}. Interpolation from this data is done during the calculation. The precision in $m_V$ is limited by the grid we created which is of the order of 10 MeV. For higher precision, the user will have to regenerate the spectra with a finer grid. By default, the Boltzmann equations are only solved up till the epoch of matter-radiation equality when the cosmic time is $t = 2\times 10^{12}$ s. This means that for a very long-lived vector boson particle $\tau \gtrsim 10^{12}$\,s, the bounds from the photodisintegration of light element abundances are conservative. While the current mass range is limited to $2m_e \leq m_V \leq 1$ GeV, we plan to extend this model to higher mass range in the future. 

To test the viability of a light vector boson parameter space, one can run the executable \texttt{decayvector} in the terminal with the command
\\\\
\texttt{./decayvector 700 1e8 1 1e-6 0.1 0.1 0.7 0.01 0.09}
\\\\
where the input parameters after \texttt{./decayvector} are
\\\\
$m_V\;\,[\mathrm{MeV}] \quad \tau\;\,[\mathrm{s}] \quad T_0\;\,[\mathrm{MeV}] \quad \left.\dfrac{n_V}{n_\gamma}\right|_{T_0} \quad \text{BR}_{ee} \quad \text{BR}_{\mu\mu} \quad \text{BR}_{\pi\pi} \quad \text{BR}_{\pi\gamma} \quad \text{BR}_{3\pi}$
\\\\
where $\text{BR}_{ee}$, $\text{BR}_{\mu\mu}$, $\text{BR}_{\pi\pi}$, $\text{BR}_{\pi\gamma}$, and $\text{BR}_{3\pi}$ denote respectively the branching ratios for $V \to e^+ e^-$, $V \to \mu^+ \mu^-$, $V \to \pi^+ \pi^-$, $V \to \pi^0 \gamma$, and $V \to \pi^0 \pi^+ \pi^-$. 
The number density of $V$ over the photon density $n_V/n_\gamma$ is related to $Y_V = n_V/s$ as follows
%%%
\begin{eqnarray}
\left.\frac{n_V}{n_\gamma}\right|_{T_0} = \frac{s(T_0)}{n_\gamma(T_0)} Y_V
= \frac{\pi^4 g_\star(T_0)}{45\zeta(3)} Y_V.
\end{eqnarray}
%%%
By default, the combined exclusion at 95\% CL for Helium-4 and deuterium considering the sum of errors Eqs.~\eqref{eq:th_error} and \eqref{eq:exp_error} in quadrature. An example of the output is as follows 
\begin{lstlisting}[language=bash, backgroundcolor=\color{white}]
Results:   Yp = 0.224415, H2/p = 0.000395, He3/p = 0.006490
Excluded by the BBN measurements at 2 sigma 
(default: He3/p not considered).
Runtime - - - 56.018600 mins - - -
\end{lstlisting}
\texttt{Yp}, \texttt{H2/p}, and \texttt{He3/p} denote respectively the Helium-4 mass fraction, $n_{\rm D}/n_{\rm H}$, and $n_{\rm He^3}/n_{\rm H}$.

%%%%%%%%%%%%%%%%%%%%%%%%%%%%%%%%%%%%%%%%
%%%%%%%%%%%%%%%%%%%%%%%%%%%%%%%%%%%%%%%%
\section{Appendix: Electromagnetic Spectra}\label{appendix-B}
The relevant decay channels of the vector boson described in section \ref{decaywidths} eventually produce a collection of photons, electrons (and positrons), and neutrinos. For studying cosmological constraints, we are mostly interested in the electromagnetic energy injected by the electrons (and positrons) and photons into the plasma in the rest frame of the decaying vector boson. The total spectra of energy are the sum of all the relevant channels
\begin{align}
    \left.\frac{dN}{dE_\alpha}\right|_V = \sum_{a} \text{BR}(V \rightarrow a) \left.\frac{dN^{(a)}}{dE_\alpha}\right|_V, \quad \alpha = e, \gamma, \quad a = e^+e^-, \mu^+\mu^-, \pi^+ \pi^-, \pi^0\gamma, \pi^0\pi^+\pi^-.
\end{align}
If the electrons (and positrons) and photons are not directly produced from the decay of the vector boson, the spectrum needs to be boosted from the ``rest frame'' of the intermediate particle to the ``lab frame'' (rest frame of the vector boson). Suppose the intermediate particle is boosted along the $x$ axis by $\gamma_I \equiv E_I/m_I$. The phase space coordinates of the electron/positron or photon in the intermediate particle's and the vector boson's rest frame are $(E^R_\alpha, \cos{\theta^R_\alpha})$ and $(E^L_\alpha, \cos{\theta^L_\alpha})$, respectively, where the angles are with respect to the $x$ axis. The partial phase space density is defined as
\begin{align}
    f(E^R_\alpha, \cos{\theta^R_\alpha}) &\equiv \dfrac{d^2N}{dE^R_\alpha d\cos{\theta_\alpha^R}}.
\end{align}
Changing the variables to the lab frame, it can be expressed as,
\begin{align}
    f\left(E^R_\alpha(E^L_\alpha, \cos{\theta^L_\alpha}), \cos{\theta^R_\alpha(E^L_\alpha, \cos{\theta^L_\alpha})}\right) &\equiv \mc J^{-1} \dfrac{d^2N}{dE^L_\alpha d\cos{\theta_\alpha^L}},\label{dNdEjac}
\end{align}
where the Jacobian is given by
\begin{align}
    \mc J = \left|
\begin{array}{cc}
 \dfrac{\partial E^R_\alpha}{\partial E^L_\alpha} & \dfrac{\partial E^R_\alpha}{\partial \cos{\theta^L_\alpha}} \\[1em]
 \dfrac{\partial \cos{\theta^R_\alpha}}{\partial E^L_\alpha} & \dfrac{\partial \cos{\theta^R_\alpha}}{\partial \cos{\theta^L_\alpha}} \\
\end{array}
\right|. 
\end{align}
Suppose the momentum $4$-vectors are 
\begin{align}
    \text{rest frame}: {p^R_\alpha}^\mu \equiv \left(E_\alpha^R, p_\alpha^R\cos{\theta_\alpha^R},p_\alpha^R\sin{\theta_\alpha^R},0\right), \\
    \text{lab frame}: {p^L_\alpha}^\mu \equiv \left(E_\alpha^L, p_\alpha^L\cos{\theta_\alpha^L},p_\alpha^L\sin{\theta_\alpha^L},0\right).
\end{align}
Their components are related by the Lorentz transformations
\begin{align}
    E^R_\alpha &= \gamma_I (E^L_\alpha - \beta_I {p^L_\alpha} \cos{\theta^L_\alpha}),\label{ER}\\
    {p^R_\alpha} \cos{\theta^R_\alpha} &= \gamma_I ({p^L_\alpha} \cos{\theta^L_\alpha}-\beta_I E^L_\alpha), \label{pRc}\\
    {p^R_\alpha} \sin{\theta^R_\alpha} &= {p^L_\alpha} \sin{\theta^L_\alpha}, \label{pRs}
\end{align}
where $\beta_I = \sqrt{1-1/\gamma_I^2}$. Writing ${p^L_\alpha} = \sqrt{{E^L_\alpha}^2 - m_\alpha^2}$ with $m_\alpha =0$ for photon and $m_\alpha = m_e$ for electron/positron, and $\sin{\theta^L_\alpha} = \sqrt{1-\cos^2{\theta^L_\alpha}}$, Eqs.~\eqref{ER}, \eqref{pRc} and \eqref{pRs} yield
\begin{align}
    E^R_\alpha &= \gamma_I\left(E^L_\alpha - \beta_I \sqrt{{E^L_\alpha}^2 - m_\alpha^2} \cos{\theta^L_\alpha}\right), \label{ER2} \\
    \cos{\theta^R_\alpha} &= \dfrac{\gamma_I \left(\sqrt{{E^L_\alpha}^2 - m_\alpha^2} \cos{\theta^L_\alpha} - \beta_I E^L_\alpha\right)}{\sqrt{\left({E^L_\alpha}^2 - m_\alpha^2\right) (-1+\cos^2{\theta^L_\alpha})+\gamma_I^2\left(\sqrt{{E^L_\alpha}^2 - m_\alpha^2} \cos{\theta^L_\alpha} - \beta_I E^L_\alpha\right)^2}}. \label{cosR}
\end{align}
From Eqs.~\eqref{ER2} and \eqref{cosR}, the Jacobian can be explicitly written in terms of the mass and energy of the intermediate particle and final radiation or electron/positron
\begin{align}
    \mc J &= \dfrac{\sqrt{{E^L_\alpha}^2 - m_\alpha^2}}{  \sqrt{\left(1-\cos^2{\theta^L_\alpha}\right) {\left({E^L_\alpha}^2 - m_\alpha^2\right)}+\gamma_I ^2 \left(\cos{\theta^L_\alpha} {\sqrt{{E^L_\alpha}^2 - m_\alpha^2}}-\beta_I  E^L_\alpha\right)^2}}. \label{jacob2}
\end{align}
For all the decay channels we will consider, $f(E^R_\alpha, \cos{\theta^R_\alpha})$ is independent of $\cos{\theta_\alpha^R}$, hence 
\begin{align}
    f(E^R_\alpha, \cos{\theta^R_\alpha}) &\equiv \dfrac{1}{2} \dfrac{dN}{dE^R_\alpha},
\end{align}
and Eq.~\eqref{dNdEjac} can be expressed as 
\begin{align}
    \dfrac{dN}{dE^L_\alpha} = \Theta\left(E_I - m_I \right) \dfrac{1}{2} \int_{-1}^{+1}d\cos{\theta_\alpha^L}\ \mc J \dfrac{dN}{dE^R_\alpha}, \label{boost}
\end{align}
where $\Theta(x) = 1$ for $x>0$ and $\Theta(x) = 0$ for $x<0$ is the Heaviside step function.

In the following we calculate the combined electron-positron spectra and photon spectrum for each of the five channels at leading order. 
\subsection{$V\rightarrow e^+ e^-$}
The leading-order electron-positron spectra is given by 
\begin{align}
    \frac{dN}{dE_e} = 2\ \delta(E_e-m_V/2).
\end{align}
Photon spectrum arises from the final state radiation (FSR) and is given by~\cite{Coogan:2019qpu, Forestell:2018txr}
\begin{align}
    \frac{dN^\text{FSR}}{dE_\gamma} &=\Theta\left(1-\frac{4 m_e^2}{m_V (m_V-2 E_\gamma)}\right) \frac{\alpha}{\pi  E_\gamma m_V  \left(m_V^2+2 m_e^2\right)\sqrt{m_V^2-4 m_e^2}} \nonumber \\
    &\times \left[ \left\{m_V^4+m_V^2 (m_V-2 E_\gamma)^2-8 m_e^2 \left(E_\gamma m_V+m_e^2\right)\right\} \log \left(\frac{1+r}{1-r}\right)\right.\nonumber\\
    &- \left.2r m_V \left\{m_V^3-2 E_\gamma m_V^2+2 m_V \left(E_\gamma^2+m_e^2\right)-4 E_\gamma m_e^2 \right\} \dfrac{}{}\!\! \right], \label{FSR}
\end{align}
where $r = \sqrt{1-\frac{4 m_e^2}{m_V (m_V-2 E_\gamma)}}$.

\subsection{$V \rightarrow \mu^+ \mu^-$}
The dominant decay process for producing electrons (and positrons) and neutrinos in this channel is muon decay $\mu^- \rightarrow e^- \bar{\nu}_e \nu_\mu$. In the muon's rest frame, the electron spectrum is given by~\cite{Coogan:2019qpu}
\begin{align}
    \frac{dN}{dE_e^R} = -\Theta\left(E_e^R - m_e\right) \frac{16}{m_\mu^4} \sqrt{{E_e^R}^2-m_e^2} \left[ 2m_e^2 - \frac{3E_e^R}{m_\mu} \left(m_e^2+m_\mu^2\right) + 4{E_e^R}^2 \right], \label{dNdEeMu}
\end{align}
and can be boosted to the lab frame by Eq.~\eqref{boost}, using Eqs.~\eqref{ER2} and \eqref{jacob2}. The positron spectrum from $\mu^+ \rightarrow e^+ \nu_e \bar{\nu}_\mu$ is exactly same.

The photon spectrum consists of an FSR part and a radiative decay $\mu^- \rightarrow e^- \bar{\nu}_e \nu_\mu \gamma$. The FSR contribution can be calculated using Eq.~\eqref{FSR} replacing $m_e$ with $m_\mu$. The radiative contribution for $\mu^-$ decay in its rest frame can be written as~\cite{Coogan:2019qpu}
\begin{align}
    \dfrac{dN^{\text{rad}}}{dE_\gamma^R} &= \frac{\alpha  ( {m_\mu-2E_\gamma^R}) }{18 \pi  {E_\gamma^R} m_\mu^4} \left[6 \left(3 m_\mu^3-4 {E_\gamma^R} m_\mu^2+16 {E_\gamma^R}^2 m_\mu-16 {E_\gamma^R}^3\right) \log \left(\frac{m_\mu (m_\mu-2 {E_\gamma^R})}{m_e^2}\right)\right.\nonumber\\
    &-\left.\left(51 m_\mu^3-46 {E_\gamma^R} m_\mu^2+202 {E_\gamma^R}^2 m_\mu-220 {E_\gamma^R}^3\right)\right] \Theta\left(\frac{m_\mu^2-m_e^2}{2m_\mu}-E_\gamma^R\right),
\end{align}
and can be boosted to the lab frame as before.

%%%%%%%%%%%%%%%%%%%%%%%%%%%%%%%%%%%%%%%%%%
\subsection{$V \rightarrow \pi^+ \pi^-$}
Monochromatic electron/positron produced from $\pi^\mp\to e^\mp \nu_e$ has a helicity suppressed branching fraction of order $BR(\pi^\mp\to e^\mp \nu_e)\sim \mathcal{O}(10^{-4})$. As a result, the dominant electron/positron spectrum originates from charged pion decays through $\pi^\mp\to \mu^\mp \nu_\mu$ mode. Muon decays almost $100\%$ of the time via electrons and the corresponding electron spectrum in the muon rest frame is given in  Eq.~\eqref{dNdEeMu}.  This spectrum can be boosted to the pion rest frame and subsequently to the lab frame using the method described above.

A combination of FSR and radiative decays contribute to the resulting photon spectrum from $V\to \pi^+\pi^-$ decay. The expression for FSR spectrum from $\pi^+\pi^-$ is given in Ref.~\cite{Coogan:2019qpu}, which has the following form: 
\begin{align}
\left.\frac{dN}{dE_\gamma}\right|_{\text{FSR}}   =  
\frac{2\alpha}{\pi E_\gamma (1-4\mu)^{3/2}} \bigg\{ &
(1-x-2\mu^2)(1-4\mu^2)\log \left[  \frac{1+(1-4\mu^2/(1-x))^{1/2}}{1-(1-4\mu^2/(1-x))^{1/2}}  \right]
\nonumber \\&
-(1-4\mu^2/(1-x))^{1/2} \left[ (1-x)(1-4\mu^2)-x^2 \right]
\bigg\},
\end{align}
where $\mu=m_\pi/m_V$ and $x=2E_\gamma/m_V$. 

The radiative decay contribution arise from pion decay to leptons~\cite{Coogan:2019qpu},
\begin{align}
\left.\frac{dN}{dE_\gamma}\right|_{\pi^+\to\ell^+\nu_\ell}    =  
\frac{\alpha\left[ f(x)+g(x)  \right]}{24\pi m_{\pi^+}f^2_\pi (r-1)^2(x-1)^2r x},
\end{align}
where $r=m^2_\ell/m^2_{\pi^+}$ and the functions $f(x)$ and $g(x)$ take the following forms: 
\begin{align}
f(x)=(r+x-1)\bigg\{&
m^2_{\pi^+}x^4(F^2_A+F^2_V)(r^2-rx+r-2(x-1)^2)
\nonumber\\&
-12\sqrt{2} f_\pi m_{\pi^+}r(x-1)x^2 \left[ F_A(r-2x+1)+xF_V \right]
\nonumber\\&
-24f^2_\pi r(x-1)\left[ 4r(x-1)+(x-2)^2  \right]
\bigg\},
\\
g(x)=12\sqrt{2}f_\pi r(x-1)^2 &\log\left(\frac{r}{1-x} \right) 
\bigg\{ m_{\pi^+}x^2 \left[F_A(x-2r)-xF_V \right]
\nonumber \\&
+\sqrt{2}f_\pi \left[2r^2-2rx-x^2+2x-2   \right]
\bigg\},
\end{align}
where the axial form factor is $F_A=0.0119$~\cite{ParticleDataGroup:2018ovx} and the vector form factor is $F_V(q^2)=0.0254(1+0.10 (1-x))$. Then, the final expression for the charged pion decay spectrum is given by~\cite{Coogan:2019qpu},
\begin{align}
\frac{dN}{dE_\gamma}(E_{\pi^+}=&m_{\pi^+})= \sum_{\ell=e,\mu} Br(\pi^+\to \ell^+\nu_\ell)\cdot  \left.\frac{dN}{dE_\gamma}\right|_{\pi^+\to \ell^+\nu_\ell} (E_{\pi^+}=m_{\pi^+}) 
\nonumber\\
&+
Br(\pi^+\to \mu^+\nu_\mu)\cdot  \left.\frac{dN}{dE_\gamma}\right|_{\mu^\pm}  (E_\mu= \frac{m_{\pi^+}^2+m^2_\mu}{2m_{\pi^+}} ). \label{Vtopipi}
\end{align}
The above formula is computed in the rest frame and need to be boosted using Eq.~\eqref{boost} to obtain the total charged pion radiative decay spectrum.

%%%%%%%%%%%%%%%%%%%%%%%%%%%%%%%%%%%%%%%%%%
\subsection{$V\to \pi^0\gamma$}
Decay of the vector boson into neutral pion channel directly produces a monochromatic photon of energy (in the lab frame) 
\begin{align}
E_\gamma \ = \ \frac{m_V}{2}\left(1-\frac{m_{\pi^0}^2}{m_V^2}\right),    
\end{align}
as well as  a pair of boosted photons from the $\pi^0$ decay, which has energy   (in the lab frame)
\begin{align}
E_0 \ = \ \frac{m_V}{2}\left(1+\frac{m_{\pi^0}^2}{m_V^2}\right).    
\end{align}
In the pion rest frame, each of these resulting photons has energy $E'=m_{\pi^0}/2$ (and a Lorentz factor $\gamma=E_0/m_{\pi^0}$ and  as before, $\beta=\sqrt{1-\gamma^{-2}}$). The full photon spectrum is then obtained from summing these different contribution yielding,  
\begin{align}
\frac{dN_{\gamma}}{dE} \ = \ \delta(E-E_{\gamma}) 
+ \frac{2}{\beta\gamma\,m_{\pi^0}}\,B(E_0) \ ,\label{pi0gamma}
\end{align}
where
\begin{align}
B(E_0)=
\left\{
\begin{array}{ccl}
1&;&E ~\in~ [(1-\beta),\,(1+\beta)]\times (E_0/2)\\
0&;&\text{otherwise}
\end{array}
\right.\nonumber .
\end{align}

%%%%%%%%%%%%%%%%%%%%%%%%%%%%%%%%%%%%%%%%%%
\subsection{$V\to \pi^0\pi^+\pi^-$}
From the decay mode $V\to \pi^0\pi^+\pi^-$, to the leading order, photons are produced through $\pi^0\to \gamma \gamma$.  The corresponding photon spectrum is given in Eq.~\eqref{pi0gamma} (the second term). Then the resulting photon spectrum from the three pion decay mode can be obtained by,
\begin{align}
    \frac{dN_{\gamma}}{dE} \ = \ \int\!dE_+\int\!dE_-\;p_{3\pi}(E_+,E_-)\,
\frac{d\hat N_{\gamma}}{dE} \ ,\label{3pion-general}
\end{align}
with
\begin{align}
\frac{d\hat N_{\gamma}}{dE}=\frac{2}{\beta\gamma\,m_{\pi^0}}\,B(E_0)\,, 
\end{align}
and 
\begin{align}
p_{3\pi}(E_+,E_-) \ = \
\frac{1}{\Gamma_{3\pi}}\,\frac{d^2\Gamma_{3\pi}}{dE_+dE_-}. \label{disctribution}   
\end{align}
Here, $p_{3\pi}$ represents the distribution of energies with which charged and neutral pions are created from $V\to \pi^0\pi^+\pi^-$ decay. Here the decay width $\Gamma_{3\pi}$ is given in Eq.~\eqref{decaywidth3pion} and $d^2\Gamma_{3\pi}/dE_+dE_-$ has the same form as Eq.~\eqref{decaywidth3pion} except the function $\mc{I}(m_V^2)$ is replaced by the integrand
of Eq.~\eqref{integral3pion}.  Moreover, the energy of the neutral pion is  $E_0 = m_V - E_+-E_-$. 

Here, we include some new effects relative to earlier literature. In addition to the direct photons produced from $\pi^0\to \gamma \gamma$, we incorporate photons that arise from internal radiative decays of $\pi^+\pi^-$. Total photon spectrum from charged pion radiative decay is already computed in Eq.~\eqref{Vtopipi}, which is then boosted to the LAB frame, as mentioned above. The resulting spectrum $(d\widetilde N_{\gamma}/dE)$ is then plugged in Eq.~\eqref{3pion-general} $(d\hat N_{\gamma}/dE\to d\widetilde N_{\gamma}/dE)$ to obtain photon spectrum from radiative decay, and  added to the direct photon contribution to acquire the total spectrum from $V\to \pi^0\pi^+\pi^-$ decay.

Following the above discussion, the leading electron/positron spectrum results from charged pion decay to muon and its subsequent decay to electron/positron. This spectrum can readily be obtained using the boosted electron/positron spectrum computed above for charged pion decay weighted by the distribution function $p_{3\pi}(E_+,E_-) $ given in Eq.~\eqref{disctribution}.

%%%%%%%%%%%%%%%%%%%%%%%%%%%%%%%%%%%%%%%%
%%%%%%%%%%%%%%%%%%%%%%%%%%%%%%%%%%%%%%%%
\bibliography{references}
\newpage
\bibliographystyle{JHEP}
\end{document}